\documentclass[useAMS, usenatbib]{mn2e}

\usepackage{graphicx}
\usepackage{lscape}
\usepackage{amsmath}
\usepackage{textcomp}
\usepackage{longtable}
\usepackage[colorlinks,linkcolor=black,anchorcolor=black,citecolor=blue]{hyperref}
\newcommand{\Ion}[2]{#1{\,\scriptsize #2}}

\title[A Search for SGPs]{A Search for Spectral Galaxy Pairs of Overlapping Galaxies based on Fuzzy Recognition}
\author[Haifeng Yang et al.]{Haifeng Yang$^{1, 2, 3}$, 
Ali Luo$^{1}$\thanks{Email: lal@nao.cas.cn},
Xiaoyan Chen$^{1}$,
Jifu Zhang$^{2}$,
Wen Hou$^{3}$,
Jianghui Cai$^{2}$,\and
Peng Wei$^{3}$,
Juanjuan Ren$^{3}$,
Xiaojie Liu$^{3}$,
Yongheng Zhao$^{1}$\\
\\
$^{1}$Key Laboratory of Optical Astronomy, National Astronomical Observatories, Chinese Academy of Sciences, Beijing 100012, China\\
$^{2}$School of Computer Science and Technology, Taiyuan University of Science and Technology, Taiyuan,030024, China\\
$^{3}$University of Chinese Academy of Sciences, Beijing 100049, China\\}
\begin{document}


\pagerange{\pageref{firstpage}--\pageref{lastpage}} \pubyear{2014}

\maketitle
\label{firstpage}

\begin{abstract}
The Spectral Galaxy Pairs (SGPs) are defined as the composite galaxy spectra which contain two independent redshift systems. These spectra are useful for studying dust properties of the foreground galaxies. In this paper, a total of 165 spectra of SGPs are mined out from Sloan Digital Sky Survey (SDSS) Data Release 9 (DR9) using the concept of membership degree from the fuzzy set theory particularly defined to be suitable for fuzzily identifying emission lines.
The spectra and images of this sample are classified according to the membership degree and their image features, respectively. Many of these 2nd redshift systems are too small or too dim to select from the SDSS images alone, making the sample a potentially unique source of information on dust effects in low-luminosity or low-surface-brightness galaxies that are under-represented in morphological pair samples.
The dust extinction of the objects with high membership degree is also estimated by Balmer decrement. Additionally, analyses for a series of spectroscopic observations of one SGP from 165 systems indicate that a newly star-forming region of our Milky Way might occur.

\end{abstract}


\section{Introduction}

The evolutionary track of a galaxy can be shaped as its merger history, which largely depends on the properties of the rich environments. Since one of the first observations \citep{dressler1980galaxy} showed that the environment may influence the properties of galaxies, numerous further identification studies have been carried out. The environmental impacts on galaxies in clusters (especially in over-dense clusters) \citep{wake2005environmental, tyler2013star}, galaxy groups \citep{wilman2008unveiling, hou2013group}, or galaxy pairs \citep{krabbe2014interaction} have always been hot issues for observations and researches. Furthermore, it has been accepted that the density on smaller scales can be the major impetus behind the galaxy evolution \citep{lewis20022df, blanton2007aspects}, and galaxy mergers may provide the most obvious mechanism. Thus, astronomers have paid a lot of attentions on galaxy pairs. From different perspectives, interacting effects of galaxy pairs were extensively analyzed in a series of works \citep{ellison2008galaxy, ellison2010galaxy, patton2011galaxy, ellison2011galaxy, scudder2012galaxy, patton2013galaxy, ellison2013galaxyb, ellison2013galaxya}, which are the most influential works in this decade. They found that there was an enhancement in the star formation rate (SFR) of galaxy pairs. And the intensity of this enhancement is mainly correlated to galaxy evolutionary phases (the enhancement in late-type galaxies is more significant than that in early-type ones, \citealp[see][]{nikolic2004star}), stellar masses relationship (major pairs are higher than others) and their separations (this enhancement increases as pair separation decreases). The latest research \citep{satyapal2014galaxy} also showed that the fraction of AGN in the pairs increases with decreasing projected separation. In general, one criteria to differentiate the interacting and non-interacting galaxy pairs is the distance (or velocity difference ${\Delta}v$ or redshift difference ${\Delta}z$) between the two galaxies. 

Further, overlapping galaxy pairs on the line of sight are very valuable materials for dust studies. Dust composes only a small fraction of the interstellar medium in galaxies, but plays an important role in understanding the fundamental cosmological parameters. These parameters are useful for determining the mass density and the cosmological constant, as well as understanding the measured star-formation rates of the higher redshift proto-galaxies in the early universe. Dust extinction and dust mass can be estimated from differential photometry in occulting pairs of galaxies, which was firstly proposed by \citet{white1992direct}, and then applied to the known pairs by using ground-based optical images \citep{domingue1999dust}, spectroscopy \citep{domingue2000seeing} and later space-based HST images \citep{keel2001seeinga, holwerda2013occulting}. The flux ratio of two Balmer emission lines from nebular (such as H${\alpha}$/H${\beta}$) can be used to determine the dust extinction as well \citep{kennicutt1992integrated}. Recently,  more pairs were found in the SDSS spectroscopic catalogue (86 pairs, \citealp[see][]{holwerda2007spiral}) and in the Galaxy Zoo project (1993 pairs, \citealp[see][]{keel2013galaxy}). The sample selection is mainly based on photometric images. It is difficult and time-consuming to search these targets in images automatically, especially for some superimposed galaxies \citep{domingue2000seeing}. Currently, there are few auto-search methods for dual-redshift system in spectra. Cross-correlation analysis results have been reported from  SDSS DR7, which can provide several redshifts candidates for every spectrum, and finally increased the data dimensions. Thus, it is more suitable for verifying multi-redshift systems. \citet{2012ApJ...753..122T} have proposed a  mathematical approach to fitting the spectral models at different redshifts, simultaneously. This method depends on both the training model and optimization strategy. Fortunately, the emission-line galaxy pairs in nearby universe may also be distinguished by aid of composite spectra, i.e. SGPs. 

In this paper, a novel SGPs search method based on fuzzy recognition is presented and especially focused on searching for `emission+emission' dual-redshift system spectra from the SDSS DR9. As a result, a total of 165 spectra of SGPs are provided, and the spectra and images of this sample are classified according to the membership degree and their image features, respectively. For the objects with high membership degree, the dust extinction is also estimated by Balmer decrement. Additionally, analyses for a series of observations of one SGP from 165 systems indicate that a newly star-forming region of our Milky Way might occur.

The paper is organized as follows. In Section 2, sample selection and auto-search method are described in detail.  A SGP list, results analysis and preliminary discussion of extinction estimation are shown in section 3. Finally, a summary of the main work is given in section 4. 

\section{Automated Search Method}
\subsection{Sample selection}
The ninth data release of the Sloan Digital Sky Survey (SDSS DR9, \citealp[see][]{ahn2012ninth}) is the first release of the spectra from the SDSS-III's Baryon Oscillation Spectroscopic Survey (BOSS), in which about 1.5 million (1, 457, 002) massive galaxies' spectra were presented. We are honoured to get these data and choose them as the initial data set of this work. To obtain  the final sample, we reduce these spectra firstly according to the following criteria. 

All spectra of galaxy from SDSS DR9 are selected as our initial data set. Here, we do not impose the subclass limit on the spectra selection although our searching is especially focused on the galaxy spectra with two groups of emission lines, and it is mainly to avoid missing the SGPs of other subclasses automatically classified by pipeline. For example, the object NO.100 in Table \ref{table4} is a SGP, while its subclass provided by the SDSS is `null'. Thus, as long as two sets of emission lines appear in a spectrum, the spectrum is kept as a candidate.  

In order to guarantee that both H$\alpha$ and H$\beta$ fall simultaneously in the SDSS spectral wavelength coverage, we limit redshift range to z ${\leq}$ 0.4. Thus, some other emission lines such as \Ion{O}{II} ${\lambda\lambda}$3727, 3730, \Ion{O}{III} ${\lambda\lambda}$4960, 5008, \Ion{N}{II} ${\lambda\lambda}$6550, 6585 could also be employed in auto-search process.

Furthermore, it is the fitting quality of the emission lines (instead of the SNR) that is used to define the membership degree, which is utilized in searching process to measure the confidence of a spectrum belonging to SGPs  (see section 2.2.1). 

\subsection{Method description}
There are many uncertain or imprecise characteristics increasing the difficulty of auto-identification. For example, not all positive signal (e.g. noise) is true emission line; not all the emission lines have peak value for various reasons such as sky subtraction error; and some emission lines are weakened by stellar absorption, such as H${\beta}$. As mentioned above, the spectral shapes of these objects vary a lot due to their different observation environments, broadening, redshifts, and  processing errors. Therefore, it is desirable to find a method which can process and analyze the uncertain data.

In mathematics, fuzzy sets \citep{zadeh1965fuzzy} are sets whose elements have degrees of membership, and now used in different areas, such as linguistics, decision-making and clustering \citep{zhang2012outlier, zhang2013interrelation}. By contrast with the classical set theory, fuzzy set theory described with the aid of a membership function valued in the real unit interval [0, 1] permits the gradual assessment of the membership of elements in a set. In our method, this idea of the gradual assessment is introduced into the line identification.

In brief, identifying emission lines in a spectrum through fuzzy recognition is the first key step of our method. And then the emission lines with the redshift are subtracted from the spectrum. Finally, the first step is repeated twice in the residual spectrum.
\subsubsection{Fuzzy recognition and membership degree}
Let pair (O, m) be a fuzzy set of SGPs where O is a galaxy spectra set and m is a mapping function (O$\rightarrow$[0, 1]). For each spectrum x $\in$O, the value m (x) is called the membership degree of x in (O, m). In other words, every element x in O has a degree m (x) belonging to SGPs.  Meanwhile, the membership degree of every  spectrum can be defined by the quality and related characteristics of some spectral lines. Here, we choose emission lines  \{H${\alpha~\lambda}$6565, H${\beta~\lambda}$4862, \Ion{O}{III} ${\lambda\lambda}$4960, 5008, \Ion{N}{II} ${\lambda\lambda}$6550, 6585, \Ion{S}{II}${\lambda\lambda}$6718, 6733\} as L, and define the parameter d by equation (1) for every emission line of each spectrum. Thus the m (x) is an average of all d in L.

\begin{equation}     
    d (factor, fit_{err}, peak) = (k_1*factor) + (k_2*fit_{err}) + (k_3*peak)
\end{equation}
where $k_\rmn{1},~k_\rmn{2},~k_\rmn{3}$ are the proportions (weights) of the three parameters respectively and sum ($k_\rmn{1},~k_\rmn{2},~k_\rmn{3}$) = 1. \\
\indent\textbf{The parameter $factor$:} The parameter $factor$ is defined as equation (2): 
\begin{equation}
f (x, y)=
\begin{cases}
0,& ~R_x\bigcap R_y +\Delta\lambda_{xy}= \Phi ~or\\ & ~coreErr_x \geqslant 99\\
1,& ~core_x \in R_y + \Delta \lambda_{xy}\\
\frac{R_y[-1]+\Delta\lambda_{xy}-R_x[0]}{coreErr_x},&  ~core_x\geqslant R_y[-1]+\Delta\lambda_{xy}\\
\frac{R_x[-1]-R_y[0]-\Delta\lambda_{xy}}{coreErr_x}, &  ~core_x \leqslant R_y[0]+\Delta\lambda_{xy}
\end{cases}
\end{equation}
where $x,~y\in L$, $core$ and $core_{err}$ is the line center and the corresponding error respectively when fitted by Gaussian function, $R = [core-core_{err}$, $core+core_{err}$] is acceptable range of the core, $R~[0]$ is the left border, $R~[-1]$ is the right one, and $\Delta \lambda$ is the difference of two lines' rest wavelength provided by SDSS. \\
\indent Assuming that y is a known line with the best fitting quality by Gaussian function, we can use $f ~(x,~y)$ as the main parameter to measure the confidence of $x$ belonging to the set L. We suppose that $\Delta  \lambda_{xy}$ = (wavelength$_y$ - wavelength$_x$)$_{rest}$ * (1+z), $H\alpha$ is a known line with best fit quality, $R\alpha$ is the core range of H$\alpha$, and $H\beta$ is a line to identify. Then $R\alpha$ + $\Delta\lambda_{\alpha\beta}$ is the theoretically acceptable wavelength range of $H\beta$. So it is certainly accepted as $H\beta$ when $core_{\beta}\in R\alpha$ + $\Delta\lambda_{\alpha\beta}$, rejected when $R{\beta} \bigcap R\alpha$ + $\Delta\lambda_{\alpha\beta}~=~\Phi$, and otherwise, accepted as a certain degree estimated by equation (2), whose value is inversely proportional to the deviation of $core_{\beta}$ from $R\alpha$ + $\Delta\lambda_{\alpha\beta}$ and is in a real unit interval [0, 1]. \\
\indent\textbf{The parameters $fit_{err}$:} In addition, the whole fitting error $fit_{err}$ also needs concern. The smaller the $fit_{err}$ is, the stronger confidence the line has. And when fit$_{err}\geqslant 99$ (99 is the default number when fitting $\chi^2 \geq 99$ ), the line should not be accepted. Then the function $fit_{err} (x)$ is,
\begin{equation}
fit_{err} (x)=
\begin{cases}
0, & ~Err_{all}\geq 99\\
1, & ~Err_{all}\leq 1\\
(99.0-Err_{all})/Err_{all}, & ~\textsf{others}
\end{cases}
\end{equation}
where $x\in L$ and $Err_{all}$ is the sum of fitting errors by Gaussian function.\\
\indent\textbf{The parameter $peak$:} Similarly, stronger emission lines are more favoured in searching. Therefore after normalizing the flux by the standard deviation, a function $peak~(x)$ is defined as
\begin{equation}
peak (x)=
\begin{cases}
1,& ~peak_{flux}\geq 10 \\
peak_{flux}/10, & ~\textsf{others}
\end{cases}
\end{equation}

\subsubsection{A Spectral Galaxy Pairs search algorithm}

The algorithm is descried as follows:
\begin{enumerate}
  \item select 1D Galaxy Spectra Set A as the initial data set, and N1, N2, m1, m2 as input parameters (\textit{N1 and N2 are thresholds of lines number ($\leq |L|$). m1 and m2 are thresholds of lines membership degree.})
  \item select a spectrum of A, traverse each emission signal in the wavelength region ($\geqslant$6000\AA)
  \item choose the strongest one as $H\alpha$ emission line candidate, and if there is no emission signal in this wavelength region turn to next spectrum (\textit{The flag of emission signal is average (flux$_{[linecore-3,~core+3]}$) $\geq$ average (flux$_{[linecore-5,~core+5]}$) })
  \item calculate the temporary redshift z1 by $H\alpha$ line, then calculate the wavelength of other lines in L according to z1, and calculate the peak flux as well as integral flux in region $\pm$5\AA~ for each line
  \item set emission lines initial number N = 0
  \item check wavelength positions of these lines. If there is an emission line on current position, do the following two steps:\\
\indent a: let N=N+1,\\
\indent b: fit this emission line by Gaussian function and quadratic polynomial, and calculate the parameter $d_{currentLine}$
  \item check the following conditions in Table \ref{table2}\\ (\textit{Average (d in L) $\geqslant$ m1, N$\geqslant$ N1, Table \ref{table1} conditions and z1$\leqslant$0.4}), if True, turn to step (viii), otherwise mask current $H\alpha$ line and turn to step (iii)
  \item subtract these emission lines ($H\alpha$ and other lines of current redshift system)
  \item search the next redshift system by using the same method as step (iii--viii) (\textit{The check conditions in step (vii) change to Average (d in L)$\geqslant$ m2, N $\geqslant$ N2, 0 $\leqslant$ z2 $\leqslant$ 0.4, and z2-z1 $\geqslant$ 0.005})\\
 \indent If success, add this spectra into SGPs candidate set B and turn to step (x); Otherwise, go to step (x) directly
  \item select next spectrum, loops steps (ii-x) until all spectra in A have been examined
  \item record Candidate Set B 
\end{enumerate}

We also compare both redshifts calculated by our method with the one provided by SDSS pipeline and find that the results have a good agreement. Besides, all the spectra in the candidate set B are further checked by eyes or various searching rules, which is discussed in the next section.

\subsubsection{Searching rules and results}
\begin{table}
 \centering
 \begin{minipage}{84mm}
  \caption{Pre-searching rules}{\label{table1}}
  \begin{tabular}{@{}ll@{}}
\hline
   Integrated flux     & Peak flux\\   
\hline
  H$\alpha_{I} \geqslant H\beta_{I}$ & H$\alpha_{P} \geqslant H\beta_{P}$\\
 \Ion{N}{II} $\lambda6550_{I} \leqslant$ \Ion{N}{II} $\lambda6585_{I}$ & \Ion{N}{II} $\lambda6550_{P}\leqslant$ \Ion{N}{II} $\lambda6585_{P}$\\
\hline
\end{tabular}
\end{minipage}
\end{table}

\begin{table}
 \centering
 \begin{minipage}{84mm}
  \caption{Results in different  thresholds}{\label{table2}}
  \begin{tabular}{@{}clllcc@{}}
  \hline
   \multicolumn{4}{c}{Conditions} & \multicolumn{2}{c}{Record Count}\\
   Table \ref{table1}\footnote{The `$\surd$' indicates that the sample is firstly reduced according to the criteria listed in Table \ref{table1}.}
   & m1 \footnote{The `m1' is membership degree threshold and in `m1' calculation process each element degree of L is calculated by equation 1 with input parameters:\{$k_1:0.4$, $k_2:0.1$, $k_3:0.5$\}. Values in this table are the empirical thresholds.}
   & m2 \footnote{The `m2' is membership degree threshold and in `m2' calculation process each element degree is calculated by equation 1 with input parameters:\{$k_1:1, k_2:0, k_3:0$\}. Values in this table are the empirical thresholds.}
   & N \footnote{The `N' is the sum of emission signal of a spectrum detected in searching process. In the experiment, N1 and N2 is set the same low threshold N. Values in this table are the empirical thresholds.}
   & auto\footnote{The record count of auto searched.} 
   & identify\footnote{The record count of spectra identified visually.} \\
 \hline
 $\surd$ & $\geqslant$0.60 & $\geqslant$0.70 & $\geqslant$3 & 504 & 44  \\
 $\surd$ & $\geqslant$0.40 & $\geqslant$0.60 & $\geqslant$3 & 858 & 108  \\
 $\surd$ & $\geqslant$0.30 & $\geqslant$0.40 & $\geqslant$3 & 6004 &142  \\
 $\surd$ & $\geqslant$0 & $\geqslant$0 & $\geqslant$3 &55092 & 165  \\
\hline
\end{tabular}
\end{minipage}
\end{table}

This algorithm can be  briefly divided into two steps: reduced by the pre-searching rules and identified by fuzzy recognition.

\textbf{Pre-searching rules.} The data set is firstly reduced according to the pre-searching rules listed in Table \ref{table1}. These rules are defined to remove those spectra that are absolutely not SGPs from the data set avoiding unnecessary line fittings. 
As seen in the Table \ref{table1}, the integrated flux and peak flux values of the emission lines are both used so that the misjudgements caused by noises or sky subtraction errors are controlled as small as possible. It should be noted that high order Balmer lines (very weak in some spectra) and some strong oxygen lines (such as \Ion{O}{II} $\lambda\lambda$3927, 3930, \Ion{O}{III} $\lambda\lambda$4960, 5008 with no certain relationship between the strength of these lines) are not chosen as reduced criteria. After two checks for every spectrum in the data set, 55092 spectra remain.

\textbf{Membership degree thresholds.} The membership degree can fuzzily measure the confidence of a spectra belonging to the SGPs. Therefore, a reasonable choice of membership degree thresholds is vitally important in the recognition process. We have tried several groups of membership degree thresholds from loose to strict and visually checked every candidate mined out by the method mentioned above. Table \ref{table2} shows the corresponding results of different thresholds. As seen from this table, along with the search conditions becoming looser from line 1 to line 4, the record count of SGPs gets larger and the confidence of these spectra in SGPs gets lower simultaneously. It is believed that the quality of new members admitted into SGPs is very poor if the conditions get much looser. 

\textbf{The limitation of redshifts difference.} To test the ability of this approach on the redshifts difference limitation, some experiments are implemented.  A test sample of 113 star-forming galaxies spectra with SNR $\geq$50 is chosen from DR9 data. Then, any two spectra, which are transformed to rest wavelength and normalized, are combined according to the mode: the redshift $ z_1 = 0,  z_2$ are random numbers generated from different intervals (column 1 of Table \ref{table5}), and $flux~=~flux_a~+~ k~*~ flux_b$ where k$\in$ [0.5, 1.5]. Thus, five groups of synthetic spectra with dual-redshift system are produced and applied in test. As seen from the results, almost all spectra can be recognized when $\Delta z \geq 0.005$, although few objects missed since the signal is too weak. The missed spectra can also be recognized when k = 1.5. However, the search method for SGPs (physically interacting pairs) of $\Delta z \leq 0.005$ still need to be further explored in follow-up works. In a word, the lower limit of $\Delta$z in the searching process is set to 0.005.

\begin{table}
 \centering
 \begin{minipage}{84mm}
  \caption{$\Delta$z limit test results}{\label{table5}}
  \begin{tabular}{@{}lccc@{}}
\hline
  The range  of $\Delta$z  & Spectra count & \multicolumn{2}{c}{Recognized success} \\
      &   & $k\in$ [0.5, 1.5]  &  k=1.5 \\
\hline
 0       $\sim$0.001        & 27 & 0 &0\\
 0.001$\sim$0.003 & 29 & 2 & 2\\
 0.003$\sim$0.005 & 23 & 5 & 9\\
 0.005$\sim$0.008 & 23 & 21 & 23\\
 0.008$\sim$0.015 & 47 & 43 &  47\\
\hline
\end{tabular}
\end{minipage}
\end{table}

\section{Results and analysis}
\subsection{SGPs sample selected from SDSS DR9}
Table \ref{table4} in Appendix lists the SGPs sample searched from SDSS DR9 by the method mentioned above. This sample includes 165 spectra which have two redshift systems. Each spectrum is assigned a serial number (shown in column 1) that will be referred to throughout this paper. 
The basic information of these objects (including SDSS ID, RA, DEC) are gathered from the Catalogue Archive Server Jobs System of SDSS DR9 and listed in this table from column 2 to column 4. 

Two redshifts (column 5, 6) are estimated by using the observed central wavelength and the rest central wavelength of the lines in the searching process. The corresponding errors are derived from the fitting errors of the line cores. These redshifts are also crossed with XCRedshift results of  SDSS DR7 (The same values of DR9 or later release are not found). The serial numbers are marked with asterisks in the upper right corner if there are at least two values in XCRedshift results close to the values given by the proposed method in this paper.

\begin{figure*} 
\centering
\includegraphics[angle=0, scale=0.5]{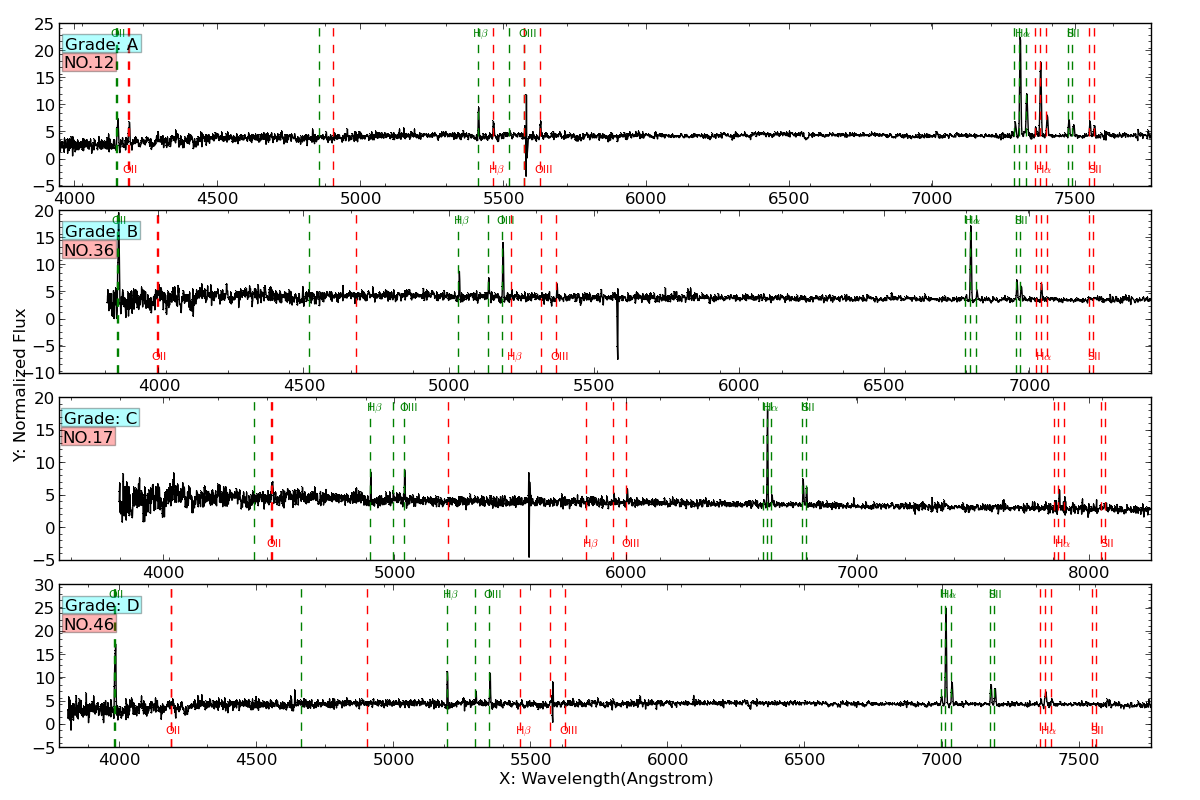}
\caption{Four examples randomly selected from each grade level (A, B, C, D) spectra. The grade level and the serial number are labelled on the top left corner of each spectrum. Two groups of common emission lines in optical band, including \{\Ion{O}{II} $\lambda\lambda$3727, 3730, H$\beta$, \Ion{O}{III} $\lambda\lambda$4960, 5008, H$\alpha$, \Ion{N}{II} $\lambda\lambda$6550, 6585, \Ion{S}{II} $\lambda\lambda$6718, 6733\}, are marked with red and green dashed lines, respectively.}
{\label{figure1}}
\end{figure*}

The confidence of spectra belonging to SGPs may vary due to different thresholds. Some spectra show very obvious characteristics in both systems, while others do not. For the latter, the lines coming from the background galaxies, can hardly be identified. In this view, these objects are divided into four grades according to different threshold values. The spectra met the conditions of line 1 in Table \ref {table2} are graded as `A', line 2 as `B', line 3 as `C', and line 4 as `D', respectively (column 7 in Table \ref{table4}). Figure \ref{figure1} shows four examples with different grades. It is noted that the grade only reflects the average fitting quality of the emission lines. And some of the weak lines may lead to the lower grade of the spectrum. Therefore some of the low-level spectra may be not reliable and require further identification.

Some objects of the sample have been observed several times. For example, the object `SDSS J131105.31+022528.3' has two spectra: No.29 and No.30, both of which appear two redshift systems. Here, for the similar objects, we only kept one spectrum with the most obvious dual-redshift system to avoid redundancy. The object `SDSS J030848.14-004659.9' (No.74), with several spectra from different observations, is a special case. It will be separately discussed later. Additionally, the basic information of No. 130 has not been found. Its Object Name is replaced by the key values string `Plate-MJD-Fiber', the RA and DEC are valued `None' in Table \ref{table4}.

\begin{figure}  
 \centering
 \includegraphics[width = 3.0in]{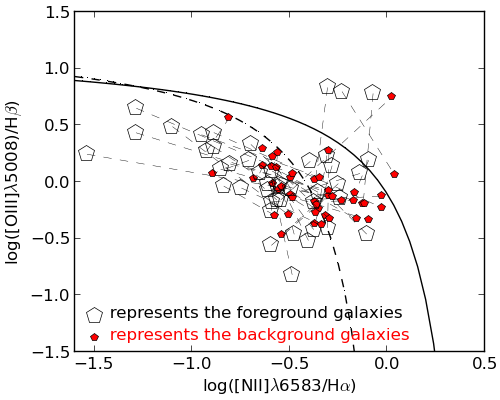}
 \caption{The emission-line ratio diagnostic diagram (i.e. BPT diagram) for 44 sources of Grade `A'. The solid curve defined by \citet{kewley2001theoretical} and dashed curve defined by \citet{kauffmann2003host} show the separation among star-forming galaxies, composite galaxies and AGNs. Two members of one spectrum are connected by a dashed line.}{\label{figure4}}
\end{figure}

\subsection{Spectra and image analysis}
\begin{figure*} 
\centering
\includegraphics[angle=0, scale=0.5]{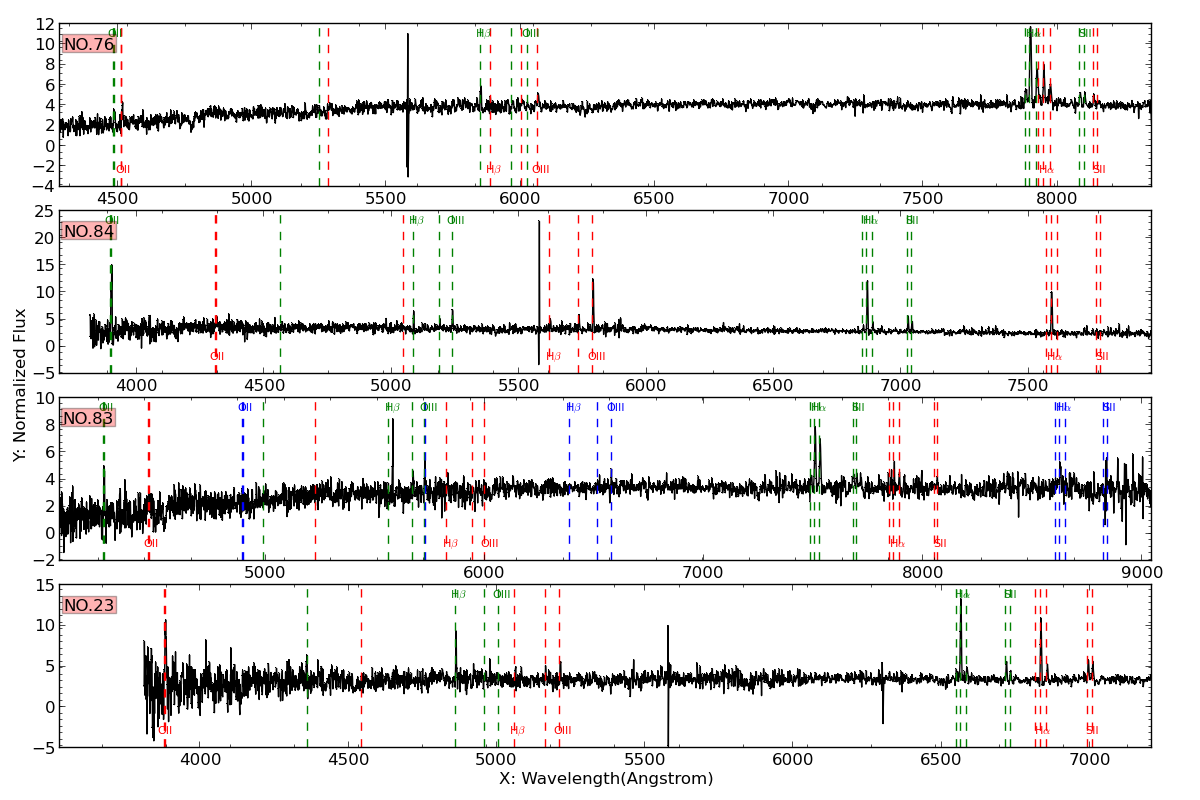}
\caption{Four special spectra: the first panel spectrum indicates that the two objects may be interacting pairs. The second panel is an interesting spectrum with strange oxygen lines. The third spectrum may have three redshift systems and emission lines of the third redshift are marked with blue dashed lines. And the last one is an example whose foreground emission are from the Milky Way. All symbols are the same as Figure \ref{figure1}. }
{\label{figure2}}
\end{figure*}

Our sample is searched based on spectra rather than the images, since those galaxies which are difficult to detect in the images can also be traced via constraints on redshifts from spectra. Figure \ref{figure1} $\&$ \ref{figure2} show some examples, both redshift systems of which are marked with dashed lines of different color. In these spectra, the most notable features are the strong emissions of H$\alpha$ band, \Ion{O}{III} $\lambda$5008 band and \Ion{O}{II} $\lambda\lambda$3727, 3729. The emission lines of most foreground galaxies are stronger than that of the background ones, and only a few spectra show the contrary appearance. 

For most spectra in this sample, the first (low) redshift systems always show obvious and strong features in the emission lines, while the second (high) redshift systems are relatively weak. Only a few SGPs show an opposite situation, which may be interesting since the foreground galaxy may be lost in the background light in the images. It is hard to clearly explain every scenes under current conditions since many factors may increase its complexity, such as dust density of foreground galaxy (opacity), varied redshifts, interstellar medium (ISM) features and different parts of the galaxy the fibre traced. Decomposition of the two members in spectra is also difficult because the luminosity contributions of the two members are not easy to determine accurately. Improving a mathematical approach, such as stellar population synthesis code  `STARLIGHT' \citep{2009A&A...495..457C}, may be a good choice. Figure \ref{figure4} shows the two members' locations of those objects graded `A' in BPT \citep{kauffmann2003host} diagram. The locations between the two members have no correlation in this diagram since that two different components without interactions may appear in one spectrum only because of superimposition in the line of sight. Most of the pairs are located in the \Ion{H}{II} region or composite region. 

\subsubsection{Some special cases} 
For the redshift difference of galaxies pairs, 0.008 is a very conservative value which would distinguish (non-)associated galaxies \citep{keel2013galaxy}. From this viewpoint, most of the SGPs in this sample are non-interacting galaxy pairs. The maximum $\Delta$z is 0.27 and the light from the background galaxy of this pair is very faint. On the contrary, the minimum value $\Delta$z $\sim$ 0.0055 (see the top panel in Figure \ref{figure2}) indicates that the two members may have weak interactions. As seen in this spectrum, the emission line \Ion{N}{II} $\lambda$6585 of the foreground galaxy and \Ion{N}{II} $\lambda$6550 of the background one are superimposed, which leads to a change of the line profile. That is why most close pairs are missed in the searching.
 
No.84 (see the second panel in Figure \ref{figure2}) is also an interesting case. The background galaxy appears to be an extremely metal-poor galaxy. And it is located on the end of left wing of the SEAGal \citep{kewley2001theoretical} in the BPT diagram, which suggests that the target is still in \Ion{H}{II} regions. In the H$\alpha$ band, \Ion{N}{II} double lines and \Ion{S}{II} double lines are almost invisible. The \Ion{O}{II} $\lambda\lambda$3727, 3730 are also very weak. However, the \Ion{O}{III} $\lambda$5007 is surprisingly strong. This feature is in stark contrast to the foreground galaxy, in which the \Ion{O}{II} double lines is stronger than \Ion{O}{III} $\lambda$5007. We speculate that it was not caused by the dust extinction effects of its foreground partner since the Balmer decrement values (H$\alpha$/H$\beta$) of the two objects are very close. The way how this spectrum forms needs further exploring.

No.83 (see the third panel in Figure \ref{figure2}) may have three redshift systems although the confidence is very low. The triplet-redshift spectra are very rare since they requires not only more than three galaxies in the line of sight, but also the emission lines triggering from overlapping areas traced by the same fibre. Even for the dual-redshift galaxy spectra, the sample (including some sources may be not true SGPs) provided in this paper is very small and occupies only 0.01 \%~ of the total galaxies spectra. In the spectrum of No.83, the first redshift system has obvious features, while the second and the third redshift system is not reliable. Especially in the third system, \Ion{S}{II} double lines are somewhat unusual, H$\beta$, \Ion{O}{II} lines do not appear in this spectrum, and \Ion{O}{III} $\lambda4364$ line is covered by the \Ion{O}{III} $\lambda5007$ line of the first system. Therefore, we speculate that the identification of H$\alpha$ band lines in the third system may be interfered by noise.

 \begin{figure*} 
\centering
\includegraphics[angle=0, scale=0.5]{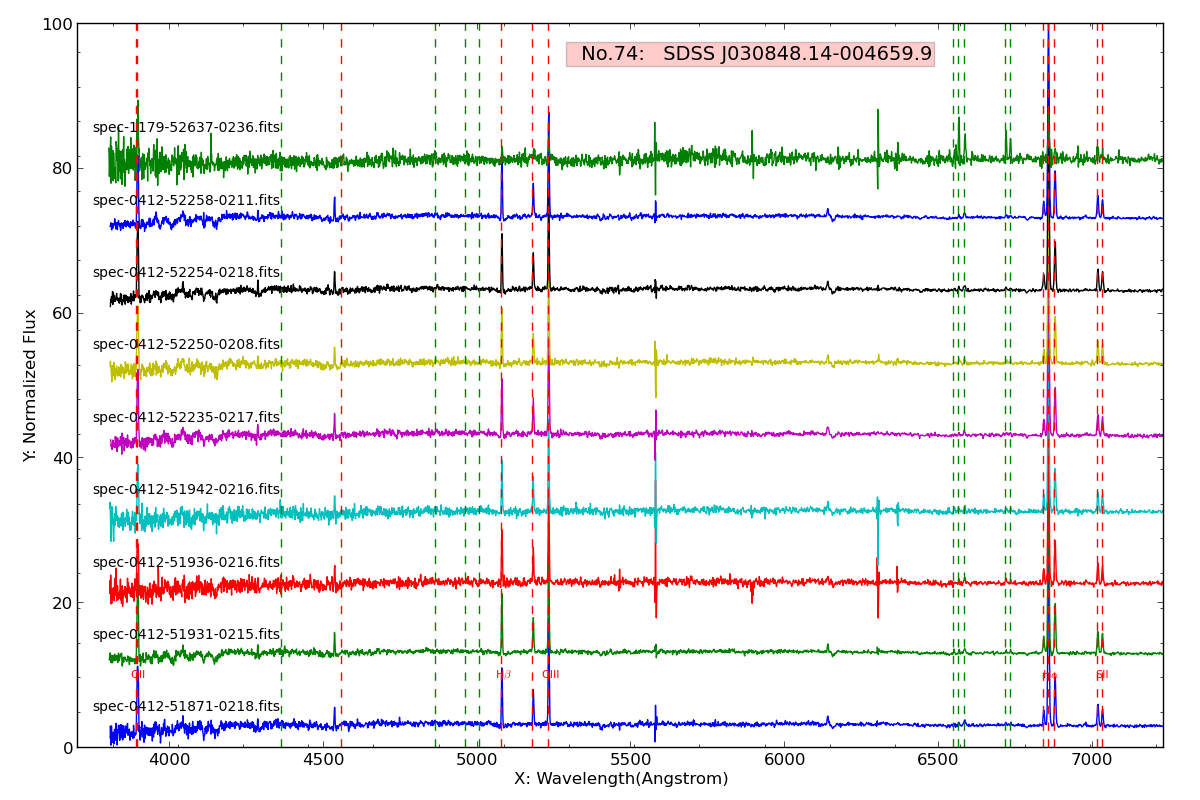}
\caption{The spectra of `SDSS J030848.14-004659.9' with nine observations sorted by modified Julian date (MJD). All spectra are scaled and shifted in one coordinate system. The fits names are marked on the top left of each spectrum. The common emission lines of two redshift systems are marked with green and red dashed lines, respectively.}
{\label{figure3}}
\end{figure*}

In this sample, there are 18 spectra (such as No.22, 23, 24, and so on) with one redshift close to 0. The bottom panel of Figure \ref{figure2} is an example (No.23). These objects are all located in sky area of Galactic anti-center (135$^\circ<$ l $<$ 225$^\circ$, -45$^\circ<$ b $<$45$^\circ$). In this direction, light from the star forming regions of the Galactic arms or local galaxies clusters (such as M31) may have more chances to sneak into fibres. Therefore, the 0-redshift systems (the redshift is very close to zero) may appear in these spectra. These spectra are useful to study the structure and evolution of our Milky Way.  Among these objects, No.74 is very special. This object has been carried out nine spectroscopic observations (shown in Figure \ref{figure3}) by SDSS, and dual-redshift system appears only in the spectrum of the latest observation . In its 0-redshift system, H$\alpha$ band has obvious emission lines while other lines are almost invisible. Moreover, the H$\alpha$ and \Ion{N}{II} $\lambda6585$ lines just coincide with the \Ion{O}{I} lines of the background galaxy. It may result in wrong recognition by pipeline. The series of spectra in Figure \ref{figure3} indicate that the star-forming region occurs in recent years.

\subsubsection{The image analysis}
\begin{figure*}
 \centering
 \includegraphics[scale=0.38]{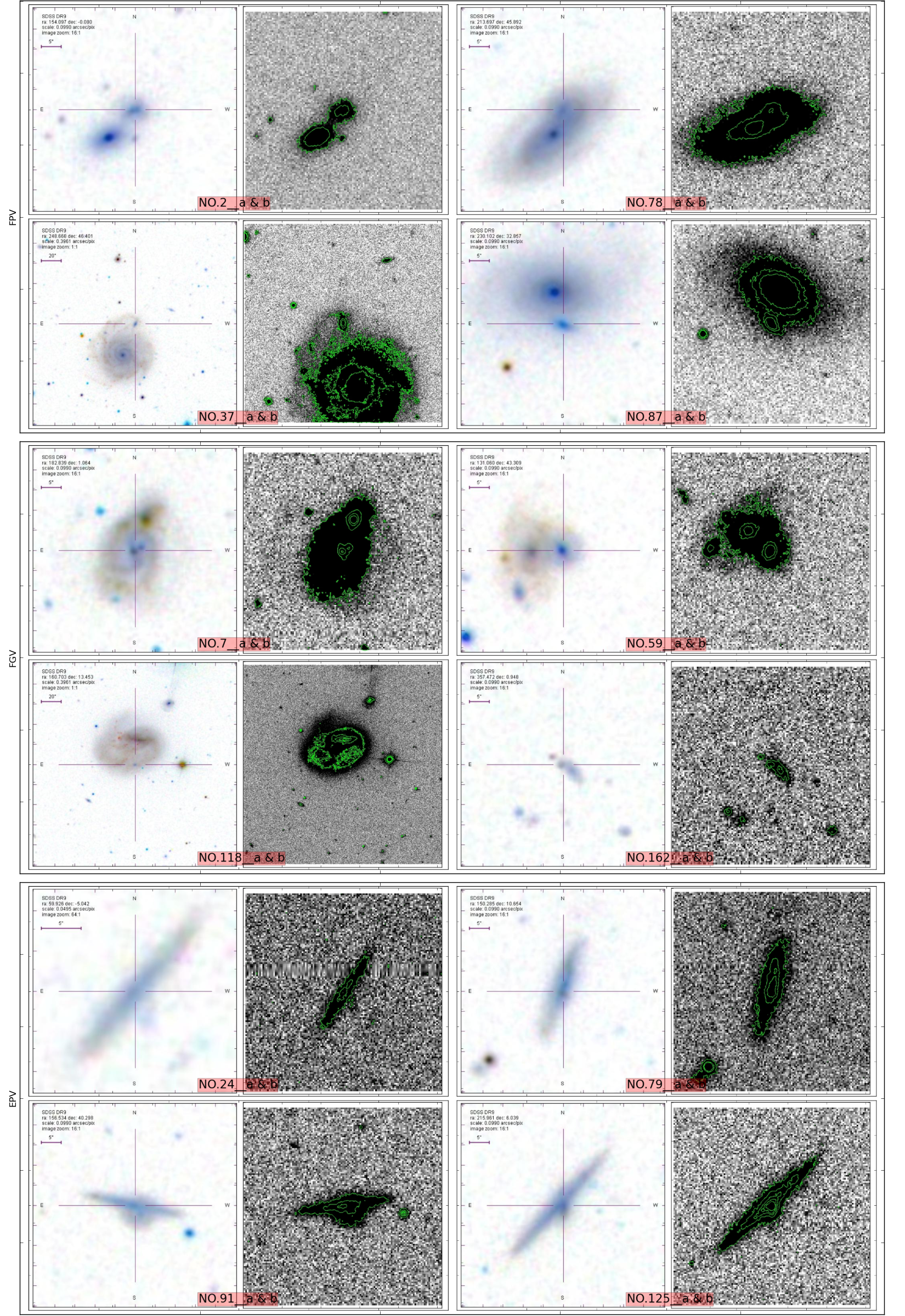}
 \caption{The examples of three subclasses (FPV, FGV, EPV) with multiple image components. The serial numbers are shown in the bottom center of every image. The scale of images is indicated  in the top left corner. The orientation is north up and east left. For every image, the left panel is a multicolour image and the right one is the corresponding contour map representation in the SDSS r-band image.} {\label{figure5}}
\end{figure*}

\begin{figure*}
 \centering
 \includegraphics[scale=0.4]{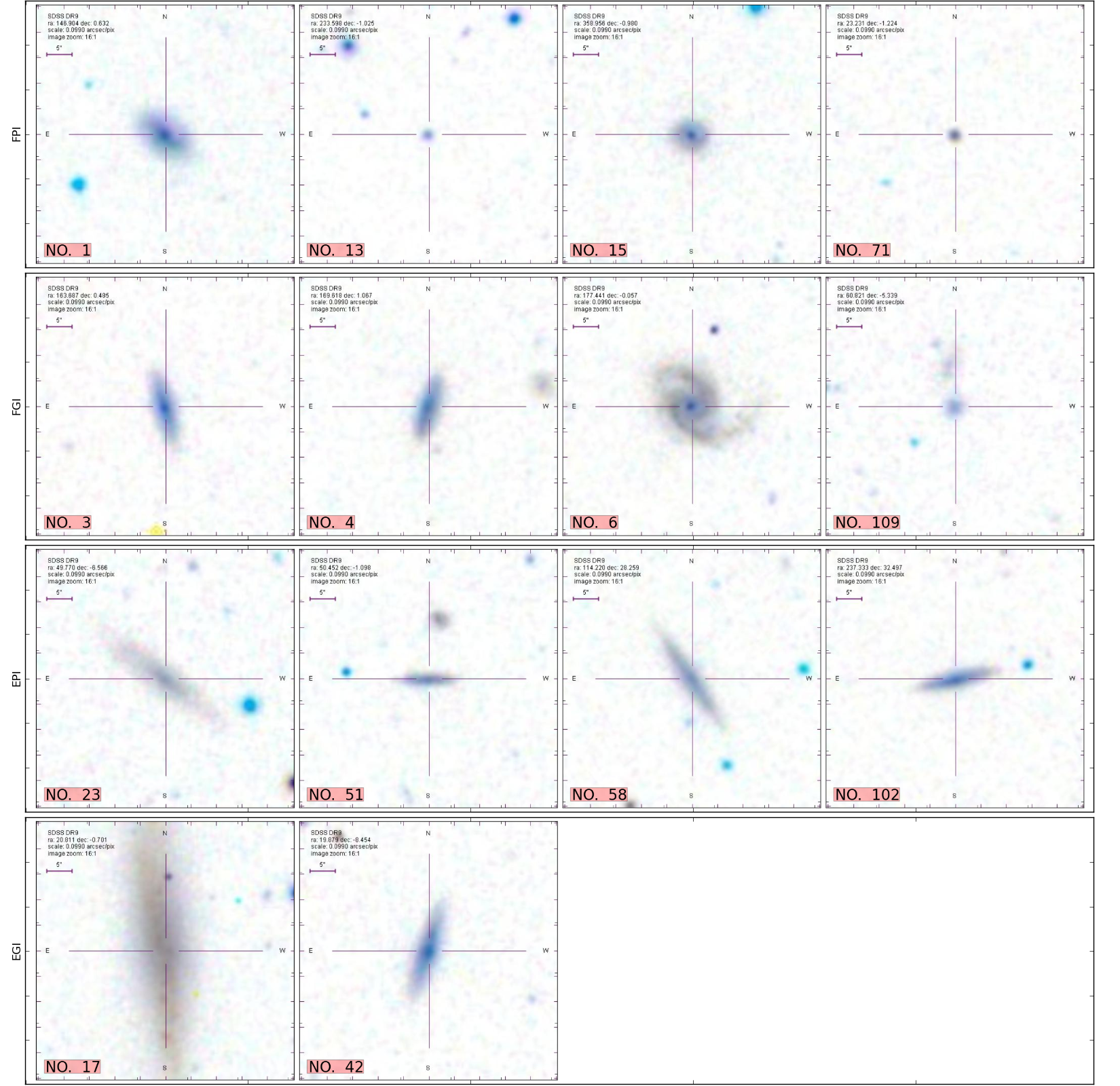}
 \caption{The examples of four subclasses (FPI, FGI, EPI, EGI) with single image component. Except that the contour map representations are absent, all symbols are the same as Figure \ref{figure5}.}
 {\label{figure6}}
\end{figure*}

In fact, the SGPs are different from the galaxy pairs. Galaxy pairs generally mean the binary systems with only two close objects. However, a SGP implies such a scene that one component exists behind its foreground partner and the emission lines are triggered in their overlapped parts. In other words, they may exist in the environment of galaxy pairs, galaxy groups, galaxy clusters, or even super galaxy clusters. To understand the scenes of their environment, photometric image is also necessary. 
So all images of this sample provided by SDSS are checked visually and divided into two basic classes, eight subclasses (see column 8 in Table \ref{table4}) according to projection features and related literature (see column 9 in Table \ref{table4}). The classification criteria of the basic class is whether the objects in one image can be distinguished (Figure \ref{figure5}) or not (Figure \ref{figure6}). The details are as follows:

\textbf{FPV} (Face-on-Pair-Visible): Two galaxies can be separated visually in the image and they are all faced on to us. The upper two rows of Figure \ref{figure5} are four examples. For the first two sources, the visual sizes of the pairs are close in the images while that of the other two sources differ considerably. The target may be a satellite galaxy of another large galaxy. 

\textbf{FGV} (Face-on-Group-Visible): They are multiple galaxy systems. Most of the galaxies in these systems are faced on to us and several objects can be separated in the images. It is noted that all sources appearing in the galaxy group, galaxy cluster or other large size structure samples in literature (see column 9 in Table \ref{table4}) are classified as `Group'. So there may be some sources showing only two galaxies in their images. The third and fourth row of Figure \ref{figure5} are four examples. 

\textbf{EPV} (Edge-on-Pair-Visible): The main difference with FPV is that these galaxies are edge-on to us. The fifth and sixth row of Figure \ref{figure5} also show four examples.

\textbf{EGV} (Edge-on-Group-Visible): The main difference with FGV is that these galaxies are edge-on to us. Unfortunately, it is very difficult to distinguish from the images. So no source in our list is classified as EGV.  

Figure \ref{figure6} shows another four subclasses examples which can not be easily separated from the images. In other words, the background galaxies are highly overlapped with the foreground ones. The main basis of this classification is from the related literature. Similar to the previous four subclasses mentioned above, they are \textbf{FPI} (Face-on-Pair-Invisible), \textbf{FGI} (Face-on-Group-Invisible), \textbf{EPI} (Edge-on-Pair-Invisible) and \textbf{EGI} (Edge-on-Group-Invisible), respectively.

\subsection{Dust extinction measurement}
 
\begin{table} 
 \centering
 \begin{minipage}{84mm}
  \caption{Dust Properties for 44 Grade `A' SGPs}{\label{table3}}
  \begin{tabular}{@{}llcccc@{}}
  \hline
   No.  & \multicolumn{2}{c}{Foreground Galaxy} & \multicolumn{2}{c}{Background Galaxy} \\
   & $H\alpha/H\beta$  & E (B-V) &$H\alpha/H\beta$  & E (B-V) \\
 \hline
 2	&	6.08$\pm$0.46	&	0.67$\pm$0.01	&	4.34$\pm$0.62	&	0.38$\pm$0.02\\
12	&	4.58$\pm$0.26	&	0.42$\pm$0.01	&	5.33$\pm$1.47	&	0.56$\pm$0.04\\
16	&	4.55$\pm$0.23	&	0.41$\pm$0.01	&	11.43$\pm$5.36	&	1.32$\pm$0.07\\
27	&	3.26$\pm$0.58	&	0.11$\pm$0.03	&	10.78$\pm$6.41	&	1.25$\pm$0.09\\
29	&	4.17$\pm$0.86	&	0.33$\pm$0.03	&	4.04$\pm$0.90	&	0.31$\pm$0.03\\
30	&	4.36$\pm$0.62	&	0.37$\pm$0.02	&	4.25$\pm$0.25	&	0.35$\pm$0.01\\
41	&	3.88$\pm$1.00	&	0.27$\pm$0.04	&	4.60$\pm$0.54	&	0.42$\pm$0.02\\
43	&	3.14$\pm$0.07	&	0.08$\pm$0.00	&	4.44$\pm$0.52	&	0.41$\pm$0.02\\
45	&	3.43$\pm$0.65	&	0.16$\pm$0.03	&	4.30$\pm$1.01	&	0.37$\pm$0.04\\
47	&	4.29$\pm$1.62	&	0.35$\pm$0.06	&	5.50$\pm$2.13	&	0.61$\pm$0.06\\
50	&	3.48$\pm$1.77	&	0.17$\pm$0.08	&	3.16$\pm$2.42	&	0.09$\pm$0.12\\
51	&	2.82$\pm$1.03	&	-0.01$\pm$0.06	&	4.81$\pm$1.13	&	0.45$\pm$0.04\\
52	&	3.50$\pm$0.33	&	0.17$\pm$0.01	&	4.66$\pm$0.39	&	0.42$\pm$0.01\\
53	&	2.98$\pm$1.51	&	0.03$\pm$0.08	&	5.09$\pm$0.88	&	0.53$\pm$0.03\\
55	&	10.51$\pm$1.35	&	1.15$\pm$0.02	&	3.82$\pm$1.88	&	0.26$\pm$0.08\\
63	&	3.99$\pm$0.58	&	0.29$\pm$0.02	&	3.26$\pm$0.24	&	0.12$\pm$0.01\\
70	&	3.74$\pm$0.26	&	0.23$\pm$0.01	&	5.78$\pm$0.53	&	0.63$\pm$0.01\\
73	&	2.71$\pm$0.41	&	-0.05$\pm$0.02	&	2.12$\pm$2.81	&	-0.27$\pm$0.20\\
76	&	4.24$\pm$1.46	&	0.37$\pm$0.05	&	7.37$\pm$7.45	&	0.88$\pm$0.15\\
83	&	4.84$\pm$1.13	&	0.47$\pm$0.04	&	1.94$\pm$1.26	&	-0.36$\pm$0.10\\
94	& 3.47$\pm$0.32	&	0.17$\pm$0.01	&	3.05$\pm$0.49	&	0.06$\pm$0.02\\
95	&	3.48$\pm$0.39	&	0.17$\pm$0.02	&	7.82$\pm$2.87	&	0.94$\pm$0.06\\
103	&	4.32$\pm$0.59	&	0.36$\pm$0.02	&	2.11$\pm$0.13	&	-0.26$\pm$0.01\\
113	&	10.55$\pm$5.79	&	1.12$\pm$0.08	&	4.56$\pm$2.07	&	0.41$\pm$0.07\\
114	&	5.17$\pm$2.05	&	0.53$\pm$0.06	&	4.92$\pm$1.51	&	0.49$\pm$0.05\\
115	&	5.00$\pm$1.01	&	0.49$\pm$0.03	&	3.98$\pm$0.30	&	0.30$\pm$0.01\\
121	&	4.37$\pm$0.77	&	0.37$\pm$0.03	&	5.66$\pm$0.94	&	0.60$\pm$0.03\\
123	&	3.98$\pm$0.76	&	0.28$\pm$0.03	&	6.67$\pm$7.60	&	0.77$\pm$0.17\\
124	&	10.08$\pm$2.87	&	1.09$\pm$0.04	&	6.44$\pm$3.01	&	0.71$\pm$0.07\\
129	&	2.48$\pm$0.21	&	-0.12$\pm$0.01	&	3.48$\pm$1.61	&	0.17$\pm$0.07\\
131	&	3.86$\pm$0.28	&	0.26$\pm$0.01	&	2.08$\pm$0.24	&	-0.28$\pm$0.02\\
132	&	2.94$\pm$0.63	&	0.02$\pm$0.03	&	3.32$\pm$2.08	&	0.13$\pm$0.10\\
135	&	4.19$\pm$1.77	&	0.33$\pm$0.06	&	2.76$\pm$2.40	&	-0.04$\pm$0.13\\
136	&	3.82$\pm$0.14	&	0.26$\pm$0.01	&	3.05$\pm$0.17	&	0.06$\pm$0.01\\
138	&	3.28$\pm$0.39	&	0.12$\pm$0.02	&	4.35$\pm$1.58	&	0.39$\pm$0.06\\
139	&	3.66$\pm$0.80	&	0.21$\pm$0.03	&	6.81$\pm$0.56	&	0.78$\pm$0.01\\
141	&	3.13$\pm$0.32	&	0.08$\pm$0.02	&	9.70$\pm$2.19	&	1.13$\pm$0.03\\
146	&	3.01$\pm$0.38	&	0.04$\pm$0.02	&	6.22$\pm$1.47	&	0.69$\pm$0.04\\
150	&	3.46$\pm$0.32	&	0.16$\pm$0.01	&	5.35$\pm$3.15	&	0.56$\pm$0.09\\
152	&	3.97$\pm$0.16	&	0.28$\pm$0.01	&	4.21$\pm$1.96	&	0.35$\pm$0.07\\
154	&	5.84$\pm$4.86	&	0.63$\pm$0.13	&	14.41$\pm$7.61	&	1.45$\pm$0.08\\
158	&	3.09$\pm$0.28	&	0.07$\pm$0.01	&	4.74$\pm$1.25	&	0.44$\pm$0.04\\
160	&	5.92$\pm$1.34	&	0.65$\pm$0.03	&	4.08$\pm$0.43	&	0.34$\pm$0.02\\
161	&	4.57$\pm$0.57	&	0.43$\pm$0.02	&	3.34$\pm$0.67	&	0.15$\pm$0.03\\
\hline
\end{tabular}
\begin{flushleft}
{\sc Notes:}\\
 1. The ratios of H$\alpha$ and H$\beta$ are integrated-flux-ratios after fitted by Gaussian function and a quadratic polynomial. E (B-V) are estimated by equation 5 and the unit is Mag. \\
 2. The corresponding errors are derived from the errors of integrated flux measurement according to the law of error propagation.\\  
\end{flushleft}
\end{minipage}
\end{table}

The effective dust extinction value can be measured successfully from pairs of overlapping galaxies in the nearby universe. This geometry of SGPs enables us to directly measure the extinction of light from the background galaxy as it passes through the foreground galaxy. The Balmer decrement can also be used to determine the extinction. The values of H$\alpha$/H$\beta>$2.86 \citep{calzetti2000dust} show extinction under the conditions normally assumed for galaxy \Ion{H}{II} regions. Due to the dependence on the measurement accuracy of emission lines, the H$\alpha$/H$\beta$ values for only the two members of those targets graded `A' are reported in Table \ref{table3}. And the H$\alpha$/H$\beta$ values are used to derive the extinction values according to equation 5 described below:
\begin{equation}
E (B-V)_{H\alpha/H\beta} = \frac{2.5}{k (H\beta)-k (H\alpha)}\log (\frac{H\alpha/H\beta}{2.86})
\end{equation}
where k (H$\alpha$) and k (H$\beta$) are taken from \citet{calzetti2000dust}.
The results are also listed in Table \ref{table3}.\\

Some extinction values may be higher than the values derived from photometric magnitude. We infer that it is reasonable. On the one hand, the spectrum is restricted by the aperture of fibres, so we can only trace a part of the galaxy (especially larger or nearer one) through the spectrum. Moreover, when we probe the emission lines radiated from \Ion{H}{II} regions, which tend to have higher dust density, we probably get a higher extinction. On the other hand, the companion component and the stellar population interfere with the measurement of integrated flux. It is difficult to decompose the members and deduct the contributions of continuum as well as stellar population simultaneously. So a Gaussian function plus a quadratic polynomial fitting are applied to the emission lines of observed spectra in our measurement. It may also lead to an overestimation of the values. Moreover, color magnitude measurement may also be affected by intergalactic medium and other members in its cluster (if so). To illustrate this result and explore the intrinsic dust distribution, the high-resolution images and the IFU spectra are needed.

The difference between two extinction values ($\Delta$E (B - V) = E (B - V )$_{bg}$ - E (B - V)$_{fg}$) can reflect dust features of the traced regions of the foreground galaxy in a clean environment. However, for most overlapping galaxy pairs, there are many objects around them, which will bias the estimation. Especially, the bias will be more severe for the galaxy pairs with larger distances. 

\section{Conclusions and discussions}
\noindent\textbf{The main works.}
An automatic search method of SGPs based on the fuzzy recognition is presented in this paper. The basic idea is to detect and identify the galaxy spectra which have two sets of emission lines from SDSS DR9. Limited by some factors such as low S/N, various broadening and other pre-processing errors, a lot of emission lines can not be identified accurately. Therefore, a membership degree is defined to measure the confidence of these lines. A SGPs sample of 165 targets is mined out by different thresholds and visual inspection. In this sample mined out by the above method, the minimum and maximum $\Delta$z is 0.0055 and 0.27 respectively, which indicates that there are nearly no interacting pairs. All spectra in our sample are divided into four grades (A, B, C, D) according to the membership degrees and those spectra graded 'A' are used for a estimation of the dust extinction. Additionally, analyses for a series of observations of one SGP from 165 systems indicate that a newly star-forming region of our Milky Way might occur.

By aid of the visual inspection of spectra and the contour lines from images, it is easy to get the relative position between these objects. In this paper, all images of our sample are classified into eight subclasses. Some images may be misclassified due to the limitation of observational capabilities and the complexity of environments around the targets. The four subclasses, with a single image component, are interesting since the foreground system may be a dwarf galaxy lost in the background light. \\

\noindent\textbf{About the method.}
The method presented in this paper is useful to measure the confidence of SGPs. The membership degree defined in previous text plays a key role in the searching process. Using high thresholds of  the membership degree will lead to the strong confidence of SGPs and a high probability of missing some SGPs with low spectral quality. On the contrary, the missed SGPs can be recovered when using low thresholds of the membership degree, and yet pseudo SGPs may be mixed into the result sample in such condition. Accordingly, the reasonable choice of membership degree thresholds is vitally important in searching process. Anyway, the searching result is reliable after trying different thresholds repeatedly.

On the other hand, there may still be some missing spectra in the searching process since the approach highly depends on the fitting quality of lines by Gaussian function. Poor fittings in some cases may be caused by weakness or deformation of the lines. In addition, this method has limitations in detecting the absorption features because some blended absorption lines could not be fitted by a single Gaussian function and the molecular bands could not be represented by Gaussian. Therefore, our method has only succeed in searching for dual-emission systems in single galaxy spectrum so far. Improving the measurement ability of the membership degree for uncertainty nature will be very valuable for those cases mentioned above. 

In the end, we hope that this paper can be a clue for follow-up observational study. First of all, the further identification observations for overlapping galaxies pairs are needed. Then, Limited by the fibre aperture, only 3" region of some targets can be traced. Thus, larger aperture telescope or IFU techniques are valuable for the studying of dust distribution. Additionally, higher resolution photometric images are also helpful to analyze environments of these targets.

\section*{Acknowledgements}
This work is supported by National Key Basic Research Program of China (Grant No. 2014CB845700), and the National Natural Science Foundation of China (Grant Nos 11233004, 61272263, 61301245).

The authors would like to thank the referee from Professor Keel, William for the  many suggestions that have helped improve the manuscript, Professor Sijiong Zhang for helpful suggestions, and Mr Jie Zheng, Miss Shuo Zhang, Miss Yanxin Guo, Miss Yu Bai for enthusiastic help and useful initial data organization and reduction. 

Funding for SDSS-III has been provided by the Alfred P. Sloan Foundation, the Participating Institutions, the National Science Foundation, and the U.S. Department of Energy Office of Science. The SDSS-III web site is http://www.sdss3.org/.

SDSS-III is managed by the Astrophysical Research Consortium for the Participating Institutions of the SDSS-III Collaboration including the University of Arizona, the Brazilian Participation Group, Brookhaven National Laboratory, University of Cambridge, Carnegie Mellon University, University of Florida, the French Participation Group, the German Participation Group, Harvard University, the Instituto de Astrofisica de Canarias, the Michigan State/Notre Dame/JINA Participation Group, Johns Hopkins University, Lawrence Berkeley National Laboratory, Max Planck Institute for Astrophysics, Max Planck Institute for Extraterrestrial Physics, New Mexico State University, New York University, Ohio State University, Pennsylvania State University, University of Portsmouth, Princeton University, the Spanish Participation Group, University of Tokyo, University of Utah, Vanderbilt University, University of Virginia, University of Washington, and Yale University.
\bibliography{reference}

\begin{thebibliography}{}

\bibitem[\protect\citeauthoryear{Ahn, Alexandroff, Prieto, Anderson, Anderton,
  Andrews, Aubourg, Bailey, Balbinot, Barnes et~al.,}{Ahn
  et~al.}{2012}]{ahn2012ninth}
Ahn C.~P.,  Alexandroff R.,  Prieto C.~A.,  Anderson S.~F.,  Anderton T.,
  Andrews B.~H.,  Aubourg {\'E}.,  Bailey S.,  Balbinot E.,  Barnes R.,
  et~al., 2012, ApJS, 203, 21

\bibitem[\protect\citeauthoryear{Allam, Tucker, Smith, Lee, Annis, Lin,
  Karachentsev \& Laubscher}{Allam et~al.}{2004}]{2004AJ....127.1883A}
Allam S.~S.,  Tucker D.~L.,  Smith J.~A.,  Lee B.~C.,  Annis J.,  Lin H.,
  Karachentsev I.~D.,    Laubscher B.~E.,  2004, AJ, 127, 1883

\bibitem[\protect\citeauthoryear{Berlind, Frieman, Weinberg, Blanton, Warren \&
  Collaboration}{Berlind et~al.}{2006}]{2006ApJS..167....1B}
Berlind A.~A.,  Frieman J.,  Weinberg D.~H.,  Blanton M.~R.,  Warren M.~S.,
  Collaboration S.,  2006, ApJS, 167, 1

\bibitem[\protect\citeauthoryear{Blanton \& Berlind}{Blanton \&
  Berlind}{2007}]{blanton2007aspects}
Blanton M.~R.,  Berlind A.~A.,  2007, AJ, 664, 791

\bibitem[\protect\citeauthoryear{Calzetti, Armus, Bohlin, Kinney, Koornneef \&
  Storchi-Bergmann}{Calzetti et~al.}{2000}]{calzetti2000dust}
Calzetti D.,  Armus L.,  Bohlin R.~C.,  Kinney A.~L.,  Koornneef J.,
  Storchi-Bergmann T.,  2000, AJ, 533, 682

\bibitem[\protect\citeauthoryear{Chen, Liang, Hammer, Zhao \& Zhong}{Chen
  et~al.}{2009}]{2009A&A...495..457C}
Chen X.~Y.,  Liang Y.~C.,  Hammer F.,  Zhao Y.~H.,    Zhong G.~H.,  2009, A\&A,
  495, 457

\bibitem[\protect\citeauthoryear{Domingue, Keel, Ryder \& {White III}}{Domingue
  et~al.}{1999}]{domingue1999dust}
Domingue D.~L.,  Keel W.~C.,  Ryder S.~D.,    {White III} R.~E.,  1999, AJ,
  118, 1542

\bibitem[\protect\citeauthoryear{Domingue, Keel \& {White III}}{Domingue
  et~al.}{2000}]{domingue2000seeing}
Domingue D.~L.,  Keel W.~C.,    {White III} R.~E.,  2000, AJ, 545, 171

\bibitem[\protect\citeauthoryear{Dressler \& Alan}{Dressler \&
  Alan}{1980}]{dressler1980galaxy}
Dressler Alan 1980, ApJ, 236, 351

\bibitem[\protect\citeauthoryear{Eke, Baugh, Cole, Frenk, Norberg, Peacock,
  Baldry, Bland-Hawthorn, Bridges, Cannon, Colless, Collins \& Couch}{Eke
  et~al.}{2004}]{2004MNRAS.348..866E}
Eke V.~R.,  Baugh C.~M.,  Cole S.,  Frenk C.~S.,  Norberg P.,  Peacock J.~A.,
  Baldry I.~K.,  Bland-Hawthorn J.,  Bridges T.,  Cannon R.,  Colless M.,
  Collins C.,    Couch W.,  2004, MNRAS, 348, 866

\bibitem[\protect\citeauthoryear{Ellison, Mendel, Scudder \& Patton}{Ellison
  et~al.}{2013a}]{ellison2013galaxyb}
Ellison S.~L.,  Mendel J.~T.,  Scudder J.~M.,    Patton D.~R.,  2013a, MNRAS,
  430, 3128

\bibitem[\protect\citeauthoryear{Ellison, Mendel, Scudder \& Patton}{Ellison
  et~al.}{2013b}]{ellison2013galaxya}
Ellison S.~L.,  Mendel J.~T.,  Scudder J.~M.,    Patton D.~R.,  2013b, MNRAS,
  435, 3627

\bibitem[\protect\citeauthoryear{Ellison, Patton, Mendel \& Scudder}{Ellison
  et~al.}{2011}]{ellison2011galaxy}
Ellison S.~L.,  Patton D.~R.,  Mendel J.~T.,    Scudder J.~M.,  2011, MNRAS,
  418, 2043

\bibitem[\protect\citeauthoryear{Ellison, Patton, Simard \&
  McConnachie}{Ellison et~al.}{2008}]{ellison2008galaxy}
Ellison S.~L.,  Patton D.~R.,  Simard L.,    McConnachie A.~W.,  2008, AJ, 135,
  1877

\bibitem[\protect\citeauthoryear{Ellison, Patton, Simard, McConnachie, Baldry
  \& Mendel}{Ellison et~al.}{2010}]{ellison2010galaxy}
Ellison S.~L.,  Patton D.~R.,  Simard L.,  McConnachie A.~W.,  Baldry I.~K.,
  Mendel J.~T.,  2010, MNRAS, 407, 1514

\bibitem[\protect\citeauthoryear{Ge, Hu, Wang, Bai \& Zhang}{Ge
  et~al.}{2012}]{2012ApJS..201...31G}
Ge J.-Q.,  Hu C.,  Wang J.-M.,  Bai J.-M.,    Zhang S.,  2012, ApJS, 201, 31

\bibitem[\protect\citeauthoryear{Holwerda, Keel \& Bolton}{Holwerda
  et~al.}{2007a}]{holwerda2007spiral}
Holwerda B.,  Keel W.,    Bolton A.,  2007a, AJ, 134, 2385

\bibitem[\protect\citeauthoryear{{Holwerda} \& {Keel}}{{Holwerda} \&
  {Keel}}{2013}]{holwerda2013occulting}
{Holwerda} B.~W.,  {Keel} W.~C.,  2013, A\&A, 556, A42

\bibitem[\protect\citeauthoryear{Holwerda, Keel \& Bolton}{Holwerda
  et~al.}{2007b}]{2007AJ....134.2385H}
Holwerda B.~W.,  Keel W.~C.,    Bolton A.,  2007b, AJ, 134, 2385

\bibitem[\protect\citeauthoryear{Hou, Parker, Balogh, McGee, Wilman, Connelly,
  Harris, Mok, Mulchaey, Bower et~al.,}{Hou et~al.}{2013}]{hou2013group}
Hou A.,  Parker L.~C.,  Balogh M.~L.,  McGee S.~L.,  Wilman D.~J.,  Connelly
  J.~L.,  Harris W.~E.,  Mok A.,  Mulchaey J.~S.,  Bower R.~G.,    et~al.,
  2013, MNRAS, 435, 1715

\bibitem[\protect\citeauthoryear{Kauffmann, Heckman, Tremonti, Brinchmann,
  Charlot, White, Ridgway, Brinkmann, Fukugita, Hall et~al.,}{Kauffmann
  et~al.}{2003}]{kauffmann2003host}
Kauffmann G.,  Heckman T.~M.,  Tremonti C.,  Brinchmann J.,  Charlot S.,  White
  S.~D.,  Ridgway S.~E.,  Brinkmann J.,  Fukugita M.,  Hall P.~B.,    et~al.,
  2003, MNRAS, 346, 1055

\bibitem[\protect\citeauthoryear{Keel, Manning, Holwerda, Mezzoprete, Lintott,
  Schawinski, Gay \& Masters}{Keel et~al.}{2013}]{keel2013galaxy}
Keel W.~C.,  Manning A.~M.,  Holwerda B.~W.,  Mezzoprete M.,  Lintott C.~J.,
  Schawinski K.,  Gay P.,    Masters K.~L.,  2013, PASP, 125, 2

\bibitem[\protect\citeauthoryear{Keel \& {White III}}{Keel \& {White
  III}}{2001}]{keel2001seeinga}
Keel W.~C.,  {White III} R.~E.,  2001, AJ, 121, 1442

\bibitem[\protect\citeauthoryear{{Kennicutt} \& {Robert}}{{Kennicutt} \&
  {Robert}}{1992}]{kennicutt1992integrated}
{Kennicutt} J.,  {Robert} C.,  1992, ApJ, 388, 310

\bibitem[\protect\citeauthoryear{Kewley, Dopita, Sutherland, Heisler \&
  Trevena}{Kewley et~al.}{2001}]{kewley2001theoretical}
Kewley L.~J.,  Dopita M.,  Sutherland R.,  Heisler C.,    Trevena J.,  2001,
  AJ, 556, 121

\bibitem[\protect\citeauthoryear{{Kopparapu}, {Hanna}, {Kalogera},
  {O'Shaughnessy}, {Gonz{\'a}lez}, {Brady} \& {Fairhurst}}{{Kopparapu}
  et~al.}{2008}]{2008ApJ...675.1459K}
{Kopparapu} R.~K.,  {Hanna} C.,  {Kalogera} V.,  {O'Shaughnessy} R.,
  {Gonz{\'a}lez} G.,  {Brady} P.~R.,    {Fairhurst} S.,  2008, ApJ, 675, 1459

\bibitem[\protect\citeauthoryear{Krabbe, Rosa, Dors, Pastoriza, Winge,
  H{\"a}gele, Cardaci \& Rodrigues}{Krabbe
  et~al.}{2014}]{krabbe2014interaction}
Krabbe A.,  Rosa D.,  Dors O.,  Pastoriza M.,  Winge C.,  H{\"a}gele G.,
  Cardaci M.,    Rodrigues I.,  2014, MNRAS, 437, 1155

\bibitem[\protect\citeauthoryear{Lewis, Balogh, {De Propris}, Couch, Bower,
  Offer, Bland-Hawthorn, Baldry, Baugh, Bridges et~al.,}{Lewis
  et~al.}{2002}]{lewis20022df}
Lewis I.,  Balogh M.,  {De Propris} R.,  Couch W.,  Bower R.,  Offer A.,
  Bland-Hawthorn J.,  Baldry I.~K.,  Baugh C.,  Bridges T.,    et~al., 2002,
  MNRAS, 334, 673

\bibitem[\protect\citeauthoryear{Liu, Shen, Strauss \& Hao}{Liu
  et~al.}{2011}]{2011ApJ...737..101L}
Liu X.,  Shen Y.,  Strauss M.~A.,    Hao L.,  2011, ApJ, 737, 101

\bibitem[\protect\citeauthoryear{McConnachie, Patton, Ellison \&
  Simard}{McConnachie et~al.}{2009}]{2009MNRAS.395..255M}
McConnachie A.~W.,  Patton D.~R.,  Ellison S.~L.,    Simard L.,  2009, MNRAS,
  395, 255

\bibitem[\protect\citeauthoryear{Marino, Plana, Rampazzo, Bianchi, Rosado,
  Bettoni, Galletta, Mazzei, Buson, Ambrocio-Cruz \& Gabbasov}{Marino
  et~al.}{2013}]{2013MNRAS.428..476M}
Marino A.,  Plana H.,  Rampazzo R.,  Bianchi L.,  Rosado M.,  Bettoni D.,
  Galletta G.,  Mazzei P.,  Buson L.,  Ambrocio-Cruz P.,    Gabbasov R.~F.,
  2013, MNRAS, 428, 476

\bibitem[\protect\citeauthoryear{Merch{\'a}n \& Zandivarez}{Merch{\'a}n \&
  Zandivarez}{2005}]{2005ApJ...630..759M}
Merch{\'a}n M.~E.,  Zandivarez A.,  2005, ApJ, 630, 759

\bibitem[\protect\citeauthoryear{Nikolic, Cullen \& Alexander}{Nikolic
  et~al.}{2004}]{nikolic2004star}
Nikolic B.,  Cullen H.,    Alexander P.,  2004, MNRAS, 355, 874

\bibitem[\protect\citeauthoryear{Patton, Ellison, Simard, McConnachie \&
  Mendel}{Patton et~al.}{2011}]{patton2011galaxy}
Patton D.~R.,  Ellison S.~L.,  Simard L.,  McConnachie A.~W.,    Mendel J.~T.,
  2011, MNRAS, 412, 591

\bibitem[\protect\citeauthoryear{{Patton}, {Torrey}, {Ellison}, {Mendel} \&
  {Scudder}}{{Patton} et~al.}{2013}]{patton2013galaxy}
{Patton} D.~R.,  {Torrey} P.,  {Ellison} S.~L.,  {Mendel} J.~T.,    {Scudder}
  J.~M.,  2013, MNRAS, 433, L59

\bibitem[\protect\citeauthoryear{Satyapal, Ellison, McAlpine, Hickox, Patton \&
  Mendel}{Satyapal et~al.}{2014}]{satyapal2014galaxy}
Satyapal S.,  Ellison S.~L.,  McAlpine W.,  Hickox R.~C.,  Patton D.~R.,
  Mendel J.~T.,  2014, MNRAS, 441, 1297

\bibitem[\protect\citeauthoryear{Scudder, Ellison, Torrey, Patton \&
  Mendel}{Scudder et~al.}{2012}]{scudder2012galaxy}
Scudder J.~M.,  Ellison S.~L.,  Torrey P.,  Patton D.~R.,    Mendel J.~T.,
  2012, MNRAS, 426, 549

\bibitem[\protect\citeauthoryear{Tsalmantza \& Hogg}{Tsalmantza \&
  Hogg}{2012}]{2012ApJ...753..122T}
Tsalmantza P.,  Hogg D.~W.,  2012, ApJ, 753, 122

\bibitem[\protect\citeauthoryear{Tyler, Rieke \& Bai}{Tyler
  et~al.}{2013}]{tyler2013star}
Tyler K.,  Rieke G.,    Bai L.,  2013, AJ, 773, 86

\bibitem[\protect\citeauthoryear{Wake, Collins, Nichol, Jones \& Burke}{Wake
  et~al.}{2005}]{wake2005environmental}
Wake D.~A.,  Collins C.~A.,  Nichol R.~C.,  Jones L.~R.,    Burke D.~J.,  2005,
  AJ, 627, 186

\bibitem[\protect\citeauthoryear{White \& Keel}{White \&
  Keel}{1992}]{white1992direct}
White R.~E.,  Keel W.~C.,  1992, Nature, 359, 129

\bibitem[\protect\citeauthoryear{Wilman, Pierini, Tyler, McGee, {Oemler Jr},
  Morris, Balogh, Bower \& Mulchaey}{Wilman et~al.}{2008}]{wilman2008unveiling}
Wilman D.,  Pierini D.,  Tyler K.,  McGee S.,  {Oemler Jr} A.,  Morris S.,
  Balogh M.,  Bower R.,    Mulchaey J.,  2008, AJ, 680, 1009

\bibitem[\protect\citeauthoryear{Zadeh}{Zadeh}{1965}]{zadeh1965fuzzy}
Zadeh L.~A.,  1965, Information and control, 8, 338

\bibitem[\protect\citeauthoryear{Zhang, Zhang, Chang \& Qin}{Zhang
  et~al.}{2014}]{zhang2012outlier}
Zhang J.,  Zhang S.,  Chang K.~H.,    Qin X.,  2014, International Journal of
  Systems Science, 45, 1170

\bibitem[\protect\citeauthoryear{Zhang, Zhao, Zhang, Yin \& Qin}{Zhang
  et~al.}{2013}]{zhang2013interrelation}
Zhang J.,  Zhao X.,  Zhang S.,  Yin S.,    Qin X.,  2013, Knowledge-Based
  Systems, 41, 77

\end{thebibliography}

\appendix
\onecolumn
\section*{Appendix : Sample List}
\begin{center}
\begin{longtable}{@{}llllccccl@{}}
\caption[]{{SGPs sample selected from SDSS DR9}\label{table4}}\\
\hline   No. & Object Name & RA & DEC & z1 & z2 & Level & Environment & Note \\  \endfirsthead

 \caption[]{-- continued from previous page}\\  \hline  No. & Object Name & RA & DEC & z1 & z2 & Grade & Environment & Note \\  \hline  \endhead
 
 \hline \\  \multicolumn{9}{r}{{Continued on next page}} \endfoot
 \endlastfoot
 
 \hline
1		& SDSS J094736.91+003755.8 & 146.90380 & 0.63221     	& 0.0628${\pm}$0.00001 & 0.1756${\pm}$0.00341 & C	& FPI &   	\\  
2$^{*}$		& SDSS J101623.29-000449.6 & 154.09699 & -0.08044	& 0.0948${\pm}$0.00000 & 0.1406${\pm}$0.00001 & A	& FPV &      	\\
3		& SDSS J105444.99+002904.7 & 163.68746 & 0.48461	& 0.1051${\pm}$0.01508 & 0.1257${\pm}$0.00005 & B     	& FGI & 1    	\\
4		& SDSS J111828.31+010400.1 & 169.61797 & 1.06670	& 0.0636${\pm}$0.00000 & 0.1776${\pm}$0.01508 & C	& FGI & 1, 2, 3	\\
5		& SDSS J111500.42-005116.0 & 168.75176 & -0.85442	& 0.1003${\pm}$0.00003 & 0.2711${\pm}$0.01508 & C	& FGV & 1	\\
6		& SDSS J114945.82-000325.3 & 177.44094 & -0.05702	& 0.0952${\pm}$0.00001 & 0.1967${\pm}$0.00097 & C	& FGI & 1	\\
7$^{*}$		& SDSS J121121.35+010348.9 & 182.83896 & 1.06361	& 0.0470${\pm}$0.00000 & 0.1248${\pm}$0.00003 & B     	& FGV &	1, 4	\\
8		& SDSS J130405.75-003508.8 & 196.02395 & -0.58576	& 0.0468${\pm}$0.00001 & 0.1465${\pm}$0.01508 & D	& FGI &	1	\\
9		& SDSS J133408.42+002840.2 & 203.53509 & 0.47788	& 0.0767${\pm}$0.00008 & 0.1590${\pm}$0.01508 & C	& FGV &	1	\\
10		& SDSS J140157.15-011108.8 & 210.48812 & -1.18586	& 0.0255${\pm}$0.00003 & 0.2711${\pm}$0.01508 & C	& FGV &	1, 2	\\
11$^{*}$	& SDSS J141546.22-011112.4 & 213.94254 & -1.18681	& 0.0499${\pm}$0.00000 & 0.1494${\pm}$0.01508 & B     	& FGV & 1, 2	\\
12$^{*}$	& SDSS J160818.74-002745.4 & 242.07809 & -0.46262	& 0.1128${\pm}$0.01508 & 0.1237${\pm}$0.00000 & A     	& FPI &		\\
13		& SDSS J153423.52-010130.3 & 233.59799 & -1.02508	& 0.1031${\pm}$0.00021 & 0.3737${\pm}$0.10000 & D	& FPI &		\\
14$^{*}$	& SDSS J233129.22+005217.1 & 352.87180 & 0.87145	& 0.1392${\pm}$0.00003 & 0.3157${\pm}$0.00005 & B     	& FPV &		\\
15		& SDSS J235549.52-005847.7 & 358.95639 & -0.97992	& 0.0748${\pm}$0.00000 & 0.1765${\pm}$0.01508 & C	& FPI & 	\\
16$^{*}$	& SDSS J010304.26+002827.3 & 15.76774  & 0.47427	& 0.0670${\pm}$0.00000 & 0.2570${\pm}$0.00003 & A     	& FPI &		\\
17$^{*}$	& SDSS J012314.54-004204.5 & 20.81058  & -0.70130	& 0.0066${\pm}$0.01508 & 0.1987${\pm}$0.00015 & C	& EGI & 	\\
18		& SDSS J013444.28+152421.9 & 23.68449  & 15.40609	& 0.0342${\pm}$0.00001 & 0.1755${\pm}$0.00662 & B     	& FPI &		\\
19		& SDSS J073926.08+392713.9 & 114.85872 & 39.45386	& 0.0615${\pm}$0.00002 & 0.2173${\pm}$0.01508 & B     	& FPI &		\\
20		& SDSS J085652.80+545749.4 & 134.22004 & 54.96374	& 0.0382${\pm}$0.00000 & 0.2711${\pm}$0.01508 & D	& FGI & 2	\\
21$^{*}$	& SDSS J092833.76+580628.8 & 142.14068 & 58.10804	& 0.1000${\pm}$0.01508 & 0.1055${\pm}$0.01508 & C	& FPI &		\\ 
22$^{*}$	& SDSS J031750.07-061134.7 & 49.45862  & -6.19297	& 0.0000${\pm}$0.01508 & 0.1708${\pm}$0.00163 & B     	& FPI &		\\
23		& SDSS J031904.90-063357.6 & 49.77042  & -6.56604	& 0.0000${\pm}$0.00001 & 0.0412${\pm}$0.00002 & B     	& EPI &		\\
24		& SDSS J035942.19-050232.3 & 59.92583  & -5.04231	& 0.0001${\pm}$0.00000 & 0.0652${\pm}$0.00000 & B     	& EPV &		\\
25		& SDSS J040708.86-054455.1 & 61.78695  & -5.74864	& 0.0001${\pm}$0.01508 & 0.1088${\pm}$0.00001 & B     	& FPI &		\\
26		& SDSS J124225.05+664807.2 & 190.60441 & 66.80203	& 0.0618${\pm}$0.00002 & 0.1745${\pm}$0.01508 & C	& FPI &		\\
27$^{*}$	& SDSS J115353.75+011914.5 & 178.47396 & 1.32068	& 0.0784${\pm}$0.01508 & 0.2305${\pm}$0.00067 & A     	& FGV &	1	\\
28		& SDSS J120600.06+023522.9 & 181.50028 & 2.58972	& 0.0683${\pm}$0.01508 & 0.1581${\pm}$0.00001 & D	& EPI &		\\
29		& SDSS J131105.31+022528.3 & 197.77220 & 2.42456	& 0.0675${\pm}$0.00000 & 0.1187${\pm}$0.01508 & A     	& FGI &	2	\\
30$^{*}$	& SDSS J131105.31+022528.3 & 197.77220 & 2.42456	& 0.0675${\pm}$0.01508 & 0.1186${\pm}$0.01508 & A     	& FGI &	2	\\
31$^{*}$	& SDSS J132120.68+032905.9 & 200.33617 & 3.48500	& 0.0823${\pm}$0.00000 & 0.2704${\pm}$0.00022 & B     	& FGI &	2	\\
32		& SDSS J100733.49+574934.3 & 151.88959 & 57.82624	& 0.1874${\pm}$0.00003 & 0.2714${\pm}$0.00010 & C	& FPI &		\\
33		& SDSS J090202.57+033525.5 & 135.51072 & 3.59044	& 0.0258${\pm}$0.00000 & 0.2714${\pm}$0.01508 & D	& FGI &	2	\\
34		& SDSS J141133.07+042854.1 & 212.88785 & 4.48168	& 0.0185${\pm}$0.00000 & 0.2185${\pm}$0.01508 & B     	& FGV &		\\
35		& SDSS J145247.81+044847.7 & 223.19921 & 4.81324	& 0.0645${\pm}$0.00001 & 0.1771${\pm}$0.01508 & C	& FGI &	2	\\
36$^{*}$	& SDSS J150630.83+043100.0 & 226.62850 & 4.51666	& 0.0353${\pm}$0.00000 & 0.0723${\pm}$0.00003 & B     	& FPI &		\\
37$^{*}$	& SDSS J163440.40+462402.4 & 248.66836 & 46.40070	& 0.0184${\pm}$0.00004 & 0.1405${\pm}$0.01508 & C	& FPV &		\\
38		& SDSS J213706.38-065950.0 & 324.27660 & -6.99722	& 0.0604${\pm}$0.00001 & 0.1765${\pm}$0.01508 & C	& FPI &		\\
39$^{*}$	& SDSS J233505.81-101625.2 & 353.77426 & -10.27369	& 0.0595${\pm}$0.00000 & 0.2106${\pm}$0.00002 & B     	& FPV &		\\
40		& SDSS J003345.57-093044.1 & 8.43987	  & -9.51226	& 0.0893${\pm}$0.00001 & 0.1628${\pm}$0.01508 & B     	& FPI &		\\
41$^{*}$	& SDSS J003959.54-084454.5 & 9.99815   & -8.74849	& 0.0942${\pm}$0.00001 & 0.1178${\pm}$0.00001 & A     	& FPV &	2	\\
42		& SDSS J011930.92-082716.1 & 19.87889  & -8.45447	& 0.0750${\pm}$0.00003 & 0.1753${\pm}$0.01508 & D	& EGI &	2	\\
43$^{*}$	& SDSS J004717.77-002233.4 & 11.82407  & -0.37584	& 0.1210${\pm}$0.00000 & 0.2062${\pm}$0.00003 & A     	& FPV &		\\
44		& SDSS J005424.39+011027.5 & 13.60168  & 1.17433	& 0.0861${\pm}$0.00001 & 0.1806${\pm}$0.01508 & B     	& FPI &		\\
45$^{*}$	& SDSS J014742.29-010806.8 & 26.92625  & -1.13524	& 0.0438${\pm}$0.00000 & 0.1346${\pm}$0.00001 & A     	& FPI &		\\
46		& SDSS J023558.17+011529.0 & 38.99244  & 1.25810	& 0.0683${\pm}$0.00000 & 0.1236${\pm}$0.00012 & D	& FPV &	4,5	\\
47$^{*}$	& SDSS J214425.44+114423.0 & 326.10600 & 11.73974	& 0.0259${\pm}$0.00001 & 0.2184${\pm}$0.00005 & A     	& FGV &	2	\\
48		& SDSS J233150.43+132827.9 & 352.96013 & 13.47441	& 0.1372${\pm}$0.00003 & 0.2714${\pm}$0.01508 & C	& EPI &		\\
49$^{*}$	& SDSS J081117.05+380000.0 & 122.82103 & 38.00000	& 0.1022${\pm}$0.00000 & 0.1834${\pm}$0.00004 & B     	& FPV &		\\
50		& SDSS J031201.75+005611.8 & 48.00735  & 0.93659	& 0.0699${\pm}$0.01508 & 0.2229${\pm}$0.01508 & A     	& FPI &		\\
51		& SDSS J032148.43-010552.7 & 50.45185  & -1.09802	& 0.0000${\pm}$0.00001 & 0.0673${\pm}$0.00000 & A     	& EPI &		\\
52$^{*}$	& SDSS J032043.66-001753.9 & 50.18198  & -0.29845	& 0.0001${\pm}$0.00001 & 0.0380${\pm}$0.00001 & A     	& EPI &		\\
53$^{*}$	& SDSS J030423.66-005823.7 & 46.09863  & -0.97329	& 0.0309${\pm}$0.00002 & 0.1915${\pm}$0.01508 & A     	& FPV &		\\
54$^{*}$	& SDSS J163627.02+384021.5 & 249.11260 & 38.67265	& 0.0353${\pm}$0.00003 & 0.2707${\pm}$0.01508 & C	& EPI &		\\
55$^{*}$	& SDSS J085805.75+410324.1 & 134.52398 & 41.05673	& 0.0883${\pm}$0.00002 & 0.1403${\pm}$0.00002 & A     	& FGV &	2	\\
56		& SDSS J090954.91+423944.4 & 137.47878 & 42.66233	& 0.0412${\pm}$0.00001 & 0.2705${\pm}$0.01508 & C	& FPI &		\\
57$^{*}$	& SDSS J113618.85+043455.9 & 174.07859 & 4.58222	& 0.0918${\pm}$0.00002 & 0.2900${\pm}$0.00013 & D	& FPI &		\\
58		& SDSS J073652.76+281534.1 & 114.21989 & 28.25949	& 0.0603${\pm}$0.00001 & 0.2710${\pm}$0.01508 & B     	& EPI &		\\
59		& SDSS J084414.50+431830.6 & 131.06047 & 43.30852	& 0.0274${\pm}$0.00004 & 0.1059${\pm}$0.00008 & B     	& FGV &	6	\\
60		& SDSS J100707.60+534604.7 & 151.78168 & 53.76803	& 0.0450${\pm}$0.00000 & 0.1877${\pm}$0.01508 & D	& FGI &	2	\\
61$^{*}$	& SDSS J101040.39+540642.1 & 152.66831 & 54.11170	& 0.0472${\pm}$0.00003 & 0.2312${\pm}$0.00002 & B     	& FPV &	2	\\
62		& SDSS J103828.47+560244.0 & 159.61869 & 56.04558	& 0.0464${\pm}$0.00001 & 0.0917${\pm}$0.00005 & B     	& FPI &		\\
63$^{*}$	& SDSS J141546.22-011112.4 & 213.94259 & -1.18682	& 0.0499${\pm}$0.00000 & 0.1493${\pm}$0.00000 & A     	& FPV &	1, 2	\\
64$^{*}$	& SDSS J132728.33+573904.8 & 201.86813 & 57.65136	& 0.0774${\pm}$0.00000 & 0.1155${\pm}$0.00004 & B     	& FPI &		\\
65		& SDSS J113541.07+484813.2 & 173.92113 & 48.80368	& 0.0367${\pm}$0.00002 & 0.1580${\pm}$0.00010 & D	& FGV &	2	\\
66$^{*}$	& SDSS J170254.76+341035.3 & 255.72818 & 34.17649	& 0.0364${\pm}$0.00001 & 0.0969${\pm}$0.00000 & C	& FGV &	2	\\
67$^{*}$	& SDSS J210607.89-004756.6 & 316.53288 & -0.79904	& 0.0347${\pm}$0.00000 & 0.1146${\pm}$0.00000 & B     	& FPI &		\\
68$^{*}$	& SDSS J223541.83-002948.6 & 338.92432 & -0.49684	& 0.0587${\pm}$0.00003 & 0.2149${\pm}$0.00013 & B     	& FPV &		\\
69		& SDSS J074619.61+275512.9 & 116.58173 & 27.92025	& 0.0757${\pm}$0.00000 & 0.3170${\pm}$0.01508 & C	& FPI &		\\
70$^{*}$	& SDSS J025450.49+010315.6 & 43.71036  & 1.05441	& 0.0437${\pm}$0.00000 & 0.1365${\pm}$0.01508 & A     	& FGV &		\\
71		& SDSS J013255.51-011325.7 & 23.23134  & -1.22384	& 0.1601${\pm}$0.00002 & 0.2841${\pm}$0.01000 & D	& FPI &		\\
72$^{*}$	& SDSS J004647.31-004214.5 & 11.69716  & -0.70403	& 0.0161${\pm}$0.00001 & 0.1867${\pm}$0.00009 & D	& FGV &		\\
73		& SDSS J033655.06-001549.0 & 54.22944  & -0.26362	& 0.0000${\pm}$0.01508 & 0.1486${\pm}$0.01508 & A     	& FPI &		\\
74		& SDSS J030848.14-004659.9 & 47.20060  & -0.78332	& 0.0001${\pm}$0.00003 & 0.0445${\pm}$0.00014 & B	& FPI &		\\
75$^{*}$	& SDSS J090601.23+365851.4 & 136.50514 & 36.98097	& 0.1070${\pm}$0.00006 & 0.2637${\pm}$0.00017 & D	& FPI &		\\
76$^{*}$	& SDSS J091636.13+364214.2 & 139.15056 & 36.70395	& 0.2034${\pm}$0.00005 & 0.2110${\pm}$0.01508 & A     	& FPI &		\\
77		& SDSS J054956.59-002613.2 & 87.48583  & -0.43703	& 0.0000${\pm}$0.01508 & 0.1076${\pm}$0.00000 & B     	& EPI &		\\
78$^{*}$	& SDSS J141447.34+455331.6 & 213.69730 & 45.89212	& 0.0406${\pm}$0.00001 & 0.0996${\pm}$0.00001 & B     	& FPV &		\\
79$^{*}$	& SDSS J100108.40+103914.3 & 150.28504 & 10.65399	& 0.0888${\pm}$0.00000 & 0.1071${\pm}$0.00006 & B     	& EPV &		\\
80$^{*}$	& SDSS J141109.08+545652.4 & 212.78781 & 54.94788	& 0.0422${\pm}$0.01508 & 0.0764${\pm}$0.00000 & B     	& FPV &	2, 5	\\
81$^{*}$	& SDSS J145123.60+512833.9 & 222.84835 & 51.47610	& 0.0353${\pm}$0.00000 & 0.1548${\pm}$0.00001 & B     	& FPI &		\\
82		& SDSS J153353.20+450929.6 & 233.47166 & 45.15823	& 0.0769${\pm}$0.00000 & 0.2062${\pm}$0.01508 & D	& EPI & 2	\\    
83		& SDSS J155019.92+445310.0 & 237.58306 & 44.88611	& 0.1437${\pm}$0.00003 & 0.1986${\pm}$0.00036 & A     	& FPI &		\\
84$^{*}$	& SDSS J164043.33+295611.6 & 250.18052 & 29.93657	& 0.0461${\pm}$0.00001 & 0.1563${\pm}$0.00002 & B     	& FPV &		\\
85$^{*}$	& SDSS J115018.78+452911.8 & 177.57832 & 45.48662	& 0.0108${\pm}$0.01508 & 0.1403${\pm}$0.00001 & B     	& FPV &		\\
86		& SDSS J130847.08+444300.7 & 197.19618 & 44.71688	& 0.0718${\pm}$0.00001 & 0.1148${\pm}$0.00006 & B     	& FPI &		\\
87$^{*}$	& SDSS J152024.36+325126.4 & 230.10153 & 32.85735	& 0.0619${\pm}$0.00001 & 0.0797${\pm}$0.00034 & B     	& FPV &	7	\\
88$^{*}$	& SDSS J144951.22+410721.9 & 222.46341 & 41.12276	& 0.0275${\pm}$0.00002 & 0.1708${\pm}$0.00006 & B     	& FPI &		\\
89$^{*}$	& SDSS J145921.13+384645.8 & 224.83806 & 38.77941	& 0.0918${\pm}$0.00001 & 0.1853${\pm}$0.00000 & B     	& FPV &	8	\\
90$^{*}$	& SDSS J161754.02+320627.4 & 244.47515 & 32.10763	& 0.0630${\pm}$0.00000 & 0.1174${\pm}$0.00031 & B     	& FPV &	5	\\
91$^{*}$	& SDSS J102608.19+401752.7 & 156.53415 & 40.29800	& 0.0642${\pm}$0.00001 & 0.1277${\pm}$0.00001 & B     	& EPV &		\\
92$^{*}$	& SDSS J004518.45+002546.6 & 11.32692  & 0.42967	& 0.0840${\pm}$0.00000 & 0.1241${\pm}$0.01508 & C	& FPI &		\\
93		& SDSS J010047.37-001136.8 & 15.19744  & -0.19363	& 0.1215${\pm}$0.00001 & 0.2212${\pm}$0.01508 & D	& EPI &		\\
94$^{*}$	& SDSS J012437.25-000911.2 & 21.15523  & -0.15309	& 0.0926${\pm}$0.00000 & 0.1372${\pm}$0.00001 & A     	& FPI &		\\
95$^{*}$	& SDSS J011759.09-000206.2 & 19.49619  & -0.03507	& 0.1054${\pm}$0.01508 & 0.2078${\pm}$0.00023 & A     	& FPI &		\\
96$^{*}$	& SDSS J011701.92-010733.1 & 19.25809  & -1.12588	& 0.1143${\pm}$0.01156 & 0.1243${\pm}$0.00014 & C	& FPI &		\\
97$^{*}$	& SDSS J234850.00-010005.3 & 357.20837 & -1.00149	& 0.2128${\pm}$0.00004 & 0.3418${\pm}$0.01508 & D	& FPI &		\\
98$^{*}$	& SDSS J003148.12+001132.3 & 7.95055   & 0.19233	& 0.0949${\pm}$0.01508 & 0.1254${\pm}$0.00000 & B     	& FPI &		\\
99$^{*}$	& SDSS J031955.18-001349.7 & 49.97998  & -0.23052	& 0.0001${\pm}$0.01508 & 0.1313${\pm}$0.00007 & B     	& FGI &	9	\\
100$^{*}$	& SDSS J165013.06+204431.2 & 252.55445 & 20.74200	& 0.0816${\pm}$0.00000 & 0.1817${\pm}$0.00002 & B     	& FPV &		\\
101$^{*}$	& SDSS J163500.59+244720.9 & 248.75246 & 24.78917	& 0.0577${\pm}$0.00000 & 0.1529${\pm}$0.00002 & B     	& FPV &		\\
102		& SDSS J154919.85+322950.9 & 237.33272 & 32.49748	& 0.0556${\pm}$0.00000 & 0.2049${\pm}$0.01508 & C	& EPI &		\\
103$^{*}$	& SDSS J093356.59+344108.6 & 143.48583 & 34.68575	& 0.0427${\pm}$0.01508 & 0.0607${\pm}$0.00000 & A     	& FGI &	3	\\
104$^{*}$	& SDSS J105831.73+115647.0 & 164.63228 & 11.94640	& 0.0887${\pm}$0.01508 & 0.1305${\pm}$0.00002 & C	& FPV &		\\
105$^{*}$	& SDSS J111624.10+110611.5 & 169.10045 & 11.10320	& 0.0199${\pm}$0.00000 & 0.0650${\pm}$0.00001 & B     	& FPI &	10	\\
106		& SDSS J113638.25+125125.2 & 174.15939 & 12.85702	& 0.1179${\pm}$0.01508 & 0.1560${\pm}$0.00067 & B     	& FPI &		\\
107$^{*}$	& SDSS J113319.23+073027.9 & 173.33014 & 7.50778	& 0.1351${\pm}$0.00001 & 0.1584${\pm}$0.00015 & B     	& FPV &		\\
108		& SDSS J040008.56-063023.4 & 60.03570  & -6.50651	& 0.0001${\pm}$0.00006 & 0.2392${\pm}$0.00003 & C	& FPI &		\\
109		& SDSS J040317.11-052020.1 & 60.82131  & -5.33893	& 0.0001${\pm}$0.00003 & 0.2166${\pm}$0.00002 & B     	& FGI &	9	\\
110		& SDSS J040705.67-060616.3 & 61.77365  & -6.10455	& 0.0001${\pm}$0.00003 & 0.2728${\pm}$0.00009 & C	& FPI &		\\
111$^{*}$	& SDSS J040804.87-061635.6 & 62.02031  & -6.27657	& 0.0001${\pm}$0.00000 & 0.2278${\pm}$0.00001 & D	& FPI &		\\
112$^{*}$	& SDSS J152411.38+291220.4 & 231.04742 & 29.20569	& 0.1131${\pm}$0.01508 & 0.1512${\pm}$0.00002 & C	& FPI &		\\
113$^{*}$	& SDSS J124315.93+131438.7 & 190.81641 & 13.24409	& 0.0029${\pm}$0.00007 & 0.0674${\pm}$0.00000 & A     	& FGV &		\\
114$^{*}$	& SDSS J135420.79+104540.8 & 208.58665 & 10.76135	& 0.1229${\pm}$0.00001 & 0.1386${\pm}$0.01508 & A     	& FPI &		\\
115$^{*}$	& SDSS J145237.94+091835.3 & 223.15813 & 9.30983	& 0.0547${\pm}$0.00001 & 0.1627${\pm}$0.00003 & A     	& FPV &		\\
116$^{*}$	& SDSS J150645.77+095133.0 & 226.69072 & 9.85917	& 0.0853${\pm}$0.01508 & 0.2094${\pm}$0.00002 & C	& FPI &		\\
117		& SDSS J151241.12+083929.5 & 228.17137 & 8.65820	& 0.0783${\pm}$0.00004 & 0.1584${\pm}$0.00009 & D	& FPI &		\\
118$^{*}$	& SDSS J104248.72+132710.8 & 160.70303 & 13.45300	& 0.0041${\pm}$0.00001 & 0.2025${\pm}$0.00003 & B     	& FGV &	11	\\
119$^{*}$	& SDSS J104456.19+135040.3 & 161.23418 & 13.84454	& 0.0376${\pm}$0.00001 & 0.0930${\pm}$0.00003 & B     	& FPV &		\\
120$^{*}$	& SDSS J110335.37+141329.1 & 165.89738 & 14.22476	& 0.0462${\pm}$0.00001 & 0.2645${\pm}$0.01508 & B     	& FPI &		\\
121$^{*}$	& SDSS J112810.02+144124.9 & 172.04178 & 14.69026	& 0.0378${\pm}$0.00001 & 0.0791${\pm}$0.00000 & A     	& FPI &	5	\\
122$^{*}$	& SDSS J085433.57+085243.1 & 133.63991 & 8.87866	& 0.0001${\pm}$0.00000 & 0.0630${\pm}$0.00001 & C	& FPI &		\\
123		& SDSS J085357.03+085401.1 & 133.48768 & 8.90032	& 0.0002${\pm}$0.00001 & 0.1708${\pm}$0.00008 & A     	& FPV &	9	\\
124$^{*}$	& SDSS J151147.25+055552.2 & 227.94689 & 5.93117	& 0.0528${\pm}$0.00000 & 0.0791${\pm}$0.00002 & A     	& FPI &		\\
125$^{*}$	& SDSS J142350.68+060219.0 & 215.96120 & 6.03861	& 0.0358${\pm}$0.01508 & 0.0557${\pm}$0.01508 & B     	& EPV &		\\
126		& SDSS J144928.24+063743.5 & 222.36770 & 6.62877	& 0.0999${\pm}$0.00000 & 0.1753${\pm}$0.01508 & B     	& FPI &		\\
127$^{*}$	& SDSS J162757.34+201030.0 & 246.98894 & 20.17503	& 0.0365${\pm}$0.00001 & 0.1516${\pm}$0.00003 & D	& EPI &		\\
128		& SDSS J072626.89+440844.1 & 111.61207 & 44.14561	& 0.0542${\pm}$0.01508 & 0.1752${\pm}$0.01508 & D	& EPI &		\\
129		& SDSS J075741.90+532435.2 & 119.42459 & 53.40979	& 0.0002${\pm}$0.00000 & 0.0429${\pm}$0.00001 & A     	& FPI &		\\
130$^{*}$	&	`1879-54478-0420'  & None      & None		& 0.0001${\pm}$0.00001 & 0.2048${\pm}$0.00007 & B     	& Non &		\\
131$^{*}$	& SDSS J101749.04+360004.2 & 154.45438 & 36.00118	& 0.0545${\pm}$0.00000 & 0.0896${\pm}$0.00002 & A     	& FPI &		\\
132$^{*}$	& SDSS J103029.76+304621.3 & 157.62404 & 30.77260	& 0.0441${\pm}$0.00000 & 0.1376${\pm}$0.00003 & A     	& FPI &		\\
133		& SDSS J111255.81+391644.1 & 168.23258 & 39.27892	& 0.0963${\pm}$0.00000 & 0.1563${\pm}$0.00006 & B     	& FPI &		\\
134		& SDSS J124858.58+365308.8 & 192.24413 & 36.88578	& 0.1517${\pm}$0.00000 & 0.1986${\pm}$0.01508 & C	& FPI &		\\
135$^{*}$	& SDSS J134741.29+382535.8 & 206.92208 & 38.42662	& 0.0091${\pm}$0.00000 & 0.2127${\pm}$0.00037 & A     	& FGV &		\\
136$^{*}$	& SDSS J105644.24+321959.9 & 164.18436 & 32.33333	& 0.1282${\pm}$0.00000 & 0.1402${\pm}$0.00001 & A     	& FPV &		\\
137$^{*}$	& SDSS J075452.04+155546.4 & 118.71685 & 15.92959	& 0.1025${\pm}$0.00004 & 0.1109${\pm}$0.00002 & B     	& FPI &		\\
138$^{*}$	& SDSS J120331.40+334112.8 & 180.88085 & 33.68691	& 0.0355${\pm}$0.00001 & 0.1780${\pm}$0.00001 & A     	& FPV &		\\
139$^{*}$	& SDSS J134951.29+281032.0 & 207.46375 & 28.17557	& 0.0554${\pm}$0.00000 & 0.1303${\pm}$0.00001 & A     	& FPV &		\\
140$^{*}$	& SDSS J135112.11+262225.2 & 207.80049 & 26.37369	& 0.0619${\pm}$0.00001 & 0.1416${\pm}$0.00010 & D	& FPV &		\\
141$^{*}$	& SDSS J112702.06+270850.4 & 171.75864 & 27.14735	& 0.0231${\pm}$0.00000 & 0.1980${\pm}$0.01508 & A     	& FPV &	5	\\
142		& SDSS J124342.12+272222.7 & 190.92552 & 27.37300	& 0.0245${\pm}$0.00000 & 0.1511${\pm}$0.00003 & B     	& FPI &		\\
143		& SDSS J094830.37+265515.6 & 147.12649 & 26.92105	& 0.0693${\pm}$0.01508 & 0.1033${\pm}$0.00058 & C	& FPI &		\\
144$^{*}$	& SDSS J101550.19+245406.0 & 153.95915 & 24.90168	& 0.0383${\pm}$0.00000 & 0.1721${\pm}$0.00030 & B     	& FPI &		\\
145$^{*}$	& SDSS J102403.67+245139.3 & 156.01533 & 24.86094	& 0.1063${\pm}$0.00001 & 0.1951${\pm}$0.00229 & B     	& FGV &		\\
146$^{*}$	& SDSS J095019.00+203259.1 & 147.57910 & 20.54981	& 0.0252${\pm}$0.00000 & 0.1008${\pm}$0.00001 & A     	& FPV &		\\
147		& SDSS J080649.19+114202.4 & 121.70497 & 11.70067	& 0.0723${\pm}$0.00000 & 0.1033${\pm}$0.01508 & D	& FPI &		\\
148		& SDSS J085107.33+150344.0 & 132.78040 & 15.06204	& 0.0699${\pm}$0.00001 & 0.2060${\pm}$0.00002 & B     	& FPV &		\\
149$^{*}$	& SDSS J090434.83+132222.6 & 136.14515 & 13.37296	& 0.0280${\pm}$0.00000 & 0.1861${\pm}$0.00004 & C	& FPI &		\\
150$^{*}$	& SDSS J112824.97+225422.7 & 172.10410 & 22.90631	& 0.0322${\pm}$0.00001 & 0.1353${\pm}$0.00011 & A     	& FPV &		\\
151$^{*}$	& SDSS J115412.76+250055.6 & 178.55322 & 25.01546	& 0.0738${\pm}$0.01508 & 0.1067${\pm}$0.01508 & B     	& FPV &		\\
152$^{*}$	& SDSS J162006.77+120851.1 & 245.02822 & 12.14753	& 0.0506${\pm}$0.01508 & 0.1805${\pm}$0.00174 & A     	& FPV &		\\
153$^{*}$	& SDSS J094914.92+132616.9 & 147.31218 & 13.43806	& 0.0406${\pm}$0.01508 & 0.2945${\pm}$0.00021 & B     	& FPV &		\\
154$^{*}$	& SDSS J101238.56+153334.9 & 153.16070 & 15.55971	& 0.0880${\pm}$0.00001 & 0.1241${\pm}$0.00005 & A     	& FPI &		\\
155$^{*}$	& SDSS J102458.65+164444.7 & 156.24440 & 16.74575	& 0.0452${\pm}$0.00001 & 0.1680${\pm}$0.00016 & C	& FPI &		\\
156		& SDSS J104132.16+171908.8 & 160.38401 & 17.31912	& 0.1227${\pm}$0.00001 & 0.1755${\pm}$0.00050 & C	& FPV &		\\
157$^{*}$	& SDSS J124453.41+193913.2 & 191.22256 & 19.65369	& 0.0645${\pm}$0.00001 & 0.2398${\pm}$0.00002 & B     	& FPV &		\\
158$^{*}$	& SDSS J135703.93+225619.5 & 209.26647 & 22.93882	& 0.0123${\pm}$0.01508 & 0.0636${\pm}$0.00002 & A     	& FPV &		\\
159		& SDSS J142249.53+214458.2 & 215.70640 & 21.74951	& 0.1430${\pm}$0.00001 & 0.1914${\pm}$0.01508 & C	& FPI &		\\
160$^{*}$	& SDSS J164054.38+272845.6 & 250.22661 & 27.47936	& 0.1033${\pm}$0.01508 & 0.2793${\pm}$0.00002 & A     	& FPI &		\\
161		& SDSS J024411.33-005021.0 & 41.04724  & -0.83919	& 0.1806${\pm}$0.00005 & 0.2396${\pm}$0.00018 & A     	& FPV &		\\
162		& SDSS J234953.32+005652.9 & 357.47218 & 0.94804	& 0.1622${\pm}$0.00001 & 0.2148${\pm}$0.00008 & B     	& FGV &	9	\\
163		& SDSS J021719.91-004555.2 & 34.33298  & -0.76533	& 0.0693${\pm}$0.00000 & 0.1497${\pm}$0.00001 & B     	& FGV &		\\
164		& SDSS J233404.08+013507.0 & 353.51702 & 1.58529	& 0.0440${\pm}$0.00000 & 0.1295${\pm}$0.01508 & B     	& FPV &		\\
165		& SDSS J093830.34+381935.2 & 144.62642 & 38.32646	& 0.1228${\pm}$0.00001 & 0.1808${\pm}$0.00024 & B     	& FPV &		\\
\hline 
\end{longtable}
\begin{flushleft}
{\sc Notes:}\\
1. `No.' is the serial number for every object and it will be referred to throughout this paper. The records are marked with asterisks in the upper right corner of serial numbers if there are at least two values in XCRedshift results close to the values identified by our method.\\
2.The two redshifts (errors) are determined by the core wavelength of best-fit emission lines. `Level' is classified according to different membership degree thresholds, `Environment' is classified by vision inspection and literature, and other columns are obtained from http://www.sdss3.org/.\\
3.The column `Note' shows that some objects appear in the sample of literature and the values are the relevant literature number: 1. \citet{2004MNRAS.348..866E}, 2. \citet{2005ApJ...630..759M}, 3. \citet{2006ApJS..167....1B}, 4. \citet{2004AJ....127.1883A}, 5. \citet{keel2013galaxy}, 6. \citet{2007AJ....134.2385H}, 7. \citet{2011ApJ...737..101L}, 8. \citet{2012ApJS..201...31G}, 9. \citet{2009MNRAS.395..255M}, 10. \citet{2008ApJ...675.1459K}, 11. \citet{2013MNRAS.428..476M}.
\end{flushleft}
\end{center}
\bsp
\label{lastpage}
\end{document}